\documentclass[aps,prd,twocolumn,superscriptaddress,groupedaddress,nofootinbib,longbibliography]{revtex4-2}
\usepackage[colorlinks=true,citecolor=blue,linkcolor=red,urlcolor=red]{hyperref}
\usepackage[sort&compress]{natbib}
\usepackage[margin=1in]{geometry}
\usepackage{romannum}
\usepackage{times}
\usepackage{setspace}
\usepackage{braket}
\usepackage{physics}
\usepackage{amsmath,amssymb,amsthm,mathtools}
\usepackage{bm}
\usepackage{simplewick}
\usepackage{tabularx}
\usepackage{comment}
\usepackage{xcolor}
\usepackage{lineno}
\begin{document}

\author{Anirudh Gundhi}
\email{anirudh.gundhi@units.it}
\affiliation{Department of Physics, University of Trieste, Strada Costiera 11, 34151 Trieste, Italy}
\affiliation{Istituto
	Nazionale di Fisica Nucleare, Trieste Section, Via Valerio 2, 34127 Trieste,
	Italy}

\author{Giorgia Infantino}
\email{giorgia.infantino@phd.units.it}
\affiliation{Department of Physics, University of Trieste, Strada Costiera 11, 34151 Trieste, Italy}
\affiliation{Istituto Nazionale di Fisica Nucleare, Trieste Section, Via Valerio 2, 34127 Trieste, Italy}	

\author{Angelo Bassi}
\email{abassi@units.it}
\affiliation{Department of Physics, University of Trieste, Strada Costiera 11, 34151 Trieste, Italy}
\affiliation{Istituto
	Nazionale di Fisica Nucleare, Trieste Section, Via Valerio 2, 34127 Trieste,
	Italy}
    
\title{Can classical theories of gravity  produce entanglement?}

\begin{abstract}
A recent paper published in Nature~\cite{AzizHowl2025} claims that quantum particles become entangled through their gravitational interaction, even when the gravitational potential is  classical. Here we show that the entanglement found by the authors is a consequence of letting particles diffuse from one object to the other. When this is not allowed, the wavefunction remains factorized. 
In fact, noticing that entanglement is relative to a chosen partition of the Hilbert space, we show that the choice of partition implicitly made by the authors is such that entanglement arises as soon as particles diffuse, even when no interaction potential is present. Therefore, it is the free evolution, not the classical gravitational potential, which acts as the entangling mechanism in the scenario considered by the authors.
\end{abstract}

\maketitle

The setup considered in~\cite{AzizHowl2025} involves two massive objects, each made of $N$ Klein-Gordon particles,  interacting gravitationally. Gravity is supposed to be {\it classical}, and the claim is that, this notwithstanding and contrary to standard wisdom, the two objects become entangled over time. 

The main result, i.e. Eq.~(\!(70)\!) of~\cite{AzizHowl2025_supp}, which is reported as Eq.~(\!(9)\!) in~\cite{AzizHowl2025},  is derived {\it before} assuming that the classical gravitational potential comes from semi-classical gravity—where the mutual gravitational attraction is described by the Schr\"odiger-Newton equation in the non relativistic limit. This means that the same conclusion should also hold if the gravitational field is that generated by an {\it external classical source}, which in principle can be assumed to be completely insensitive to the dynamics of the particles. That an external classical source can entangle two otherwise non-interacting systems looks counter-intuitive. 

\begin{figure}[t!]
    \centering
    \includegraphics[width=0.8\linewidth]{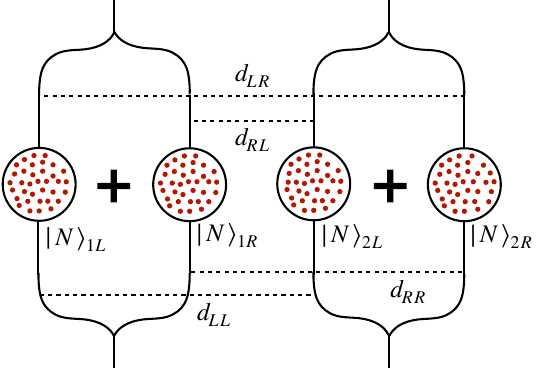}
    \caption{Objects 1 and 2, depicted by spheres comprising $N$ particles (red), are prepared  in spatial superpositions   $(\bm{X}_{1L},\bm{X}_{1R})$ and $(\bm{X}_{2L},\bm{X}_{2R})$ respectively, where $\bm{X}_{1L}$, $\bm{X}_{1R}$, $\bm{X}_{2L}$ and $\bm{X}_{2R}$ denote four different non-overlapping spatial locations.}
    \label{fig:setup}
\end{figure}

We re-examine this claim. Consider the initial state as in~\cite{AzizHowl2025}
\begin{align} \label{eq:initial}
\ket{\Psi(0)}  = \frac{1}{2}\left(\ket{N}_{1L}+\ket{N}_{1R} \right)\otimes\left(\ket{N}_{2L}+\ket{N}_{2R}\right),
\end{align}
describing each of the two objects (1 and 2) prepared in a spatial superposition, as depicted in Fig.~\ref{fig:setup}. In the QFT language, the state can be re-written as follows:
\begin{align} \label{eq:initial2}
\ket{\Psi(0)} = \frac{1}{2N!}\sum_{m=L,R} (A^\dagger_{1m})^N \sum_{k=L,R} (A^\dagger_{2k})^N|0\rangle;
\end{align}
the operators $A^\dagger_{ai}$ create a particle within the spheres $\theta_{ai}({\bf x}) = \theta(R - |{\bf x} - {\bm X}_{ai}|)$ of radius $R$ centered at locations ${\bm X}_{ai}$. See also Ref.~\cite{AzizHowl2025_supp}, and the discussion in Sec.~\ref{Sec:TransitionAmplitudes} of the supplementary material (SM).


In Sec.~\ref{sec:NoEntanglement} of the SM we show that, for a number-conserving Hamiltonian quadratic in the field operators, as in the present case, the state at time $t$ takes the form:
\begin{align} \label{eq:final}
\ket{\Psi(t)} = \frac{1}{2N!}\sum_{m=L,R} (A^\dagger_{1m}(t))^N \sum_{k=L,R} (A^\dagger_{2k}(t))^N|0\rangle,
\end{align}
where ${A}^\dagger_{ai}(t)$ now creates a particle at $\tilde\theta_{ai}(t)$, which  is $\theta_{ai}$ evolved at time $t$ under the one-particle dynamics of the problem under consideration.

The state in Eq.~\eqref{eq:final} retains the same structure  as that of Eq.~\eqref{eq:initial}, meaning that the two objects evolve independently from each other. One would quite naturally conclude that $\ket{\Psi(t)}$ is not entangled. In Refs.~\cite{Struyve2026, Diosi_GravityEntanglement}, a similar conclusion has been shown to hold also when the gravitational potential is supposed to be that coming from semiclassical gravity, i.e. the Schr\"odinger-Newton equation. 

Yet, the authors of Ref.~\cite{AzizHowl2025} claim to find entanglement. They do so by computing the probability of finding the two objects, after some time, in one of the four locations $\ket{N}_{1i} \ket{N}_{2j}$ with $i,j = L,R$, after the objects interact with the (external) gravitational potential.  The state encoding these four alternatives is:
\begin{align} \label{eq:fin_pro}
|\tilde\Psi(t)\rangle
& = \frac{1}{2N!}[
\beta_{LL} \hat A_{1L}^{\dagger N} \hat A_{2L}^{\dagger N} |0\rangle
+
\beta_{LR} \hat A_{1L}^{\dagger N} \hat A_{2R}^{\dagger N}|0\rangle
\nonumber\\
&+
\beta_{RL} \hat A_{1R}^{\dagger N} \hat A_{2L}^{\dagger N}|0\rangle
+
\beta_{RR} \hat A_{1R}^{\dagger N} \hat A_{2R}^{\dagger N}|0\rangle],
\end{align}
with $A_{ai}^{\dagger N} = (A_{ai}^\dagger)^N$ and the coefficients $\beta_{ij}$ carry the time dependence  (cf. Eq.~(\!(12)\!) of~\cite{AzizHowl2025_supp}; here we have neglected the spin). Note that $|\tilde\Psi(t)\rangle \neq |\Psi(t)\rangle$ of Eq.~\eqref{eq:final}, since the free evolution allows the objects to leak outside the four original regions. In other words, $|\tilde\Psi(t)\rangle$ is a projection of $|\Psi(t)\rangle$ onto the subspace generated by the four alternatives $\ket{N}_{1i} \ket{N}_{2j}$.

The coefficients $\beta_{ij}$ are
 the transition amplitudes\begin{align}\label{eq:transition}
\beta_{ij}(t) = 2 \times \prescript{}{1i}{\bra{N}}\prescript{}{2j}{\bra{N}}\hat{U}(t)|\Psi(0)\rangle;
\end{align}
if the associated $2 \times 2$ matrix has rank one, then these coefficients satisfy $\beta_{ij}(t) = a_{1i}(t) b_{2j}(t)$ and the state $|\tilde\Psi(t)\rangle$ is factorized. If the rank is two, the state is entangled. The authors  show, perturbatively at 4$^{\rm{th}}$ order, that the second case holds.

Sec.~\ref{Sec:TransitionAmplitudes} of the SM shows that the analysis of Ref.~\cite{AzizHowl2025} is incomplete and partly inconsistent; nevertheless it is true that the correct matrix $\beta = [\beta_{ij}]$, at 4$^{\text{th}}$ order, in general has rank two, and thus the wavefunction cannot be factorized. However, the non-factorized form of the wavefunction occurs because particles are allowed to diffuse from one object to the other, in contradiction to the physical scenario considered by the authors; see for example the discussion around Eq.~(\!(38)\!) in Ref.~\cite{AzizHowl2025_supp}. If diffusion were truly suppressed, the wavefunction would remain factorized. 
 
The SM explains in detail why the entanglement considered by the authors originates  from the free evolution. Here we show why and how this is occurs, by resorting to a simplified example, which captures the main features of the situation considered in Ref.~\cite{AzizHowl2025}. This analysis will further reconcile this result with what was stated before, in connection with Eq.~\eqref{eq:final}. 

\begin{figure}[t!]
    \centering
    \includegraphics[width=0.45\textwidth]{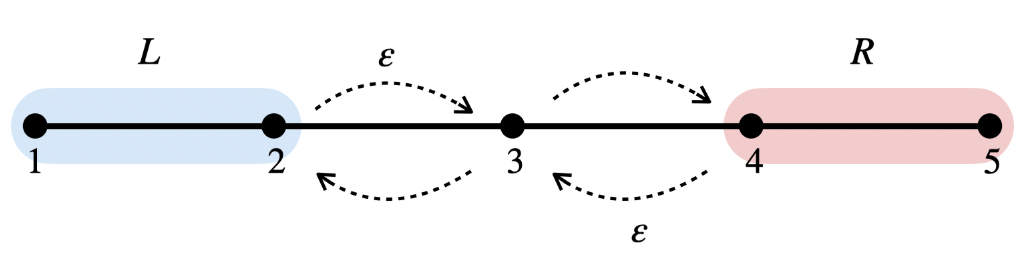}
    \caption{The lattice section highlighted in blue, including sites $1$ and $2$, corresponds to the left region, while the one highlighted in red corresponds to the right region. At time $t>0$, two bosons hop over lattice sites, from region left to right and vice versa, and lose their correspondence with the spatial regions (left or right) they belonged to at $t=0$.}\label{fig:lattice}
\end{figure}

Consider  a one-dimensional lattice consisting of five sites 1 to 5, as depicted in Fig.~\ref{fig:lattice}, and  two free bosons, not subjected to any potential; initially, one is in the superposition of sitting in sites 1 and 2, the other in the superposition of  sitting  in 4 and 5. In the QFT language, let $\hat a_i^\dagger$ and $\hat a_i$ be the bosonic creation and annihilation operators respectively, associated with the site $i$, which satisfy the canonical commutation relations. The initial state, mimicking the situation in Eq.~\eqref{eq:initial2} for $N=1$, is:
\begin{equation} \label{eq:in_sim}
\ket{\Psi(0)} = \frac{1}{2}
(\hat a_1^\dagger+\hat a_2^\dagger)
(\hat a_4^\dagger+\hat a_5^\dagger)|0\rangle.
\end{equation}

The coefficient matrix associated with the four possibilities $((1,4),(1,5),(2,4),(2,5))$, modulo the common factor $1/2$, is:
\begin{equation}
\beta(0) =
\begin{pmatrix}
1&1\\
1&1
\end{pmatrix},
\end{equation}
which has rank one, meaning that the coefficients are of the form $\beta(0)_{ij} = a(0)_i b(0)_j$. The initial state is factorized.

Over time, the two particles hop between nearest-neighbor sites---this is the analog of the free evolution in space. The second-quantized Hamiltonian is
\begin{equation}
\hat H = J\sum_{n=1}^{4}
\left(
\hat a_{n+1}^\dagger \hat a_n
+
\hat a_n^\dagger \hat a_{n+1}
\right),
\end{equation}
where $J$ is the hopping amplitude. 
The state at time $t$ is trivially: 
\begin{equation} \label{eq:fin_sim}
\ket{\Psi(t)} = \frac{1}{2}
(\hat a_1^\dagger(t) +\hat a_2^\dagger(t))
(\hat a_4^\dagger(t) +\hat a_5^\dagger(t))|0\rangle,
\end{equation}
where $\hat a^\dagger_i(t) = \hat U(t) \hat a^\dagger_i U^\dagger(t)$, with $\hat U(t) = e^{-i \hat H t/\hbar}$, creates a freely evolved boson. The situation is structurally the same as in Eq.~\eqref{eq:final} of the previous setup, except that now there is no classical potential. It is not much of a surprise, since the two particles do not interact and evolve independently.

Let us introduce the dimensionless  parameter $\epsilon=Jt/\hbar$; we  consider the weak-diffusion regime $\epsilon\ll1$, and solve the dynamics perturbatively, obtaining:
\begin{equation}
\ket{\Psi(t)} = \frac{1}{2}
\Big(
\sum_{i=1}^{5}\ell_i \hat a_i^\dagger
\Big)
\Big(
\sum_{j=1}^{5}\ell_{6-j} \hat a_j^\dagger
\Big)
|0\rangle +{\mathcal O}(\epsilon^5).
\end{equation}
The coefficients $\ell_j$ are listed in Sec.~\ref{sec:Lattice} of the SM. The number of hoppings required to reach a specific site determines the leading order in $\epsilon$ at which the various coefficients become non-zero: $\ell_1,\ell_2={\mathcal O}(1)$, $\ell_3={\mathcal O}(\epsilon)$, $\ell_4={\mathcal O}(\epsilon^2)$, and $\ell_5={\mathcal O}(\epsilon^3)$, which matches the fact that the particles mainly remain in their original locations and only weakly diffuse elsewhere.

Following the reasoning of \cite{AzizHowl2025}, we now retain only the sector in which one boson is found in $L$  and the other boson is found in $R$. In doing so, the third site is factored out when considering the final state. The four relevant Fock states are $\hat a_1^\dagger \hat a_4^\dagger|0\rangle$, $\hat a_1^\dagger \hat a_5^\dagger|0\rangle$, $\hat a_2^\dagger \hat a_4^\dagger|0\rangle$, and $\hat a_2^\dagger \hat a_5^\dagger|0\rangle$. The unnormalized projected state (analogous to~\eqref{eq:fin_pro}) is
\begin{align} \label{eq:finpro}
|\tilde\Psi(t)\rangle  = & \frac{1}{2} [
\beta_{14} \hat a_1^\dagger \hat a_4^\dagger|0\rangle
+
\beta_{15} \hat a_1^\dagger \hat a_5^\dagger|0\rangle
\nonumber\\
&+
\beta_{24} \hat a_2^\dagger \hat a_4^\dagger|0\rangle
+
\beta_{25} \hat a_2^\dagger \hat a_5^\dagger|0\rangle],
\end{align}
where
\begin{align}
\beta_{14} & =  \ell_1\ell_2+\ell_5\ell_4, \qquad
\beta_{15} =  \ell_1^2+\ell_5^2, \nonumber \\
\beta_{24} & =   \ell_2^2+\ell_4^2, \qquad
\beta_{25} =  \ell_2\ell_1+\ell_4\ell_5. \label{eq:coeff_article}
\end{align}
The coefficient matrix is therefore
\begin{equation}
\beta(\epsilon) =
\begin{pmatrix}
\beta_{14}&\beta_{15}\\
\beta_{24}&\beta_{25}
\end{pmatrix},
\end{equation}
whose determinant is
\begin{equation}
\det \beta = - ( \ell_1\ell_4 - \ell_2\ell_5)^2 = -\frac{\epsilon^4}{4} +{\mathcal O}(\epsilon^5).
\end{equation}
Thus, for any small but nonzero $\epsilon$, the rank of the matrix $\beta(\epsilon)$ is 2, meaning that the coefficients cannot be factorized as $\beta(\epsilon)_{ij} = a(\epsilon)_i b(\epsilon)_j$. One then concludes that the state in Eq.~\eqref{eq:finpro} is entangled.  This conclusion holds {\it without including any interaction, either classical or quantum.}

It is worthwhile noting that the structure of the coefficients in Eq.~\eqref{eq:coeff_article} is the same as that derived for the problem considered in~\cite{AzizHowl2025} for $N=1$ (see Secs.~\ref{Sec:TransitionAmplitudes} and~\ref{Sec:FirstQuantization} of SM), further showing that the example here is structurally equivalent to that considered in~\cite{AzizHowl2025}. 

The question, again, is how to reconcile two apparently contradictory results. 
The answer relies on the fact that {\it entanglement is relative to the partitioning of the Hilbert space} \cite{Zanardi2001TPS,Zanardi2004TPS}. As a simple illustration, consider
\begin{equation}
|\psi\rangle
=
\frac{\hat a^\dagger+\hat b^\dagger}{\sqrt{2}}|0\rangle
=
\frac{|1\rangle_a|0\rangle_b+|0\rangle_a|1\rangle_b}{\sqrt{2}}.
\end{equation}
This state is entangled with respect to
$\mathcal F=\mathcal F_a\otimes\mathcal F_b$. Defining instead
\begin{equation}
\hat c^\dagger
=
\frac{\hat a^\dagger+\hat b^\dagger}{\sqrt{2}},
\qquad
\hat d^\dagger
=
\frac{\hat a^\dagger-\hat b^\dagger}{\sqrt{2}},
\end{equation}
the same state becomes
\begin{equation}
|\psi\rangle
=
\hat c^\dagger|0\rangle
=
|1\rangle_c|0\rangle_d,
\end{equation}
and is therefore factorized with respect to
$\mathcal F=\mathcal F_c\otimes\mathcal F_d$.

Coming back to our case, the state $|\Psi(t)\rangle$ as expressed in Eq.~\eqref{eq:fin_sim} is {\it factorized} with respect to the evolved modes $\hat a_L^\dagger(t) = \hat a_1^\dagger(t) + \hat a_2^\dagger(t)$ and $\hat a_R^\dagger(t) = \hat a_4^\dagger(t) + \hat a_5^\dagger(t)$, i.e. with respect to the partition ${\mathcal P}_1$ of the Fock space ${\mathcal F}$, according to which ${\mathcal F} = {\mathcal F_{a_L(t)}} \otimes {\mathcal F_{a_R(t)}} \otimes {\mathcal F_{\perp(t)}}$, where ${\mathcal F_{\perp(t)}}$ is the Fock space generated by the remaining orthogonal one-particle modes. This is a direct consequence of the fact that the two initially occupied modes evolve independently under the quadratic, noninteracting Hamiltonian. 

The authors of~\cite{AzizHowl2025} are, instead, interested in assessing the entangling properties with respect to a different partition ${\mathcal P}_2$, specifically  $\mathcal F = \mathcal F_L\otimes\mathcal F_3\otimes\mathcal F_R$. While the initial state $|\Psi(0)\rangle$ is factorised with respect to both partitions, at later times the hopping allows the particles to diffuse from one side to the other, making them lose correspondence with their original sites. Because of this, for generic nonzero times, the state becomes {\it entangled} with respect to the  partition ${\mathcal P}_2$, while remaining factorized with respect to the  partition ${\mathcal P}_1$. This occurs because the free evolution $\hat U(t)$ does not factorize with respect to ${\mathcal P}_2$: $\hat U(t) \neq \hat U_L(t) \otimes  \hat U_3(t) \otimes \hat U_R(t)$, because of the hopping terms, thus generating entanglement with respect to that partition.

The entanglement with respect to ${\mathcal P}_2$ {\it is generated by the free evolution}. A classical interaction mimicking the classical potential considered in~\cite{AzizHowl2025},
\begin{equation}
\hat H_{\text{int}} = \sum_{i=1}^5 \Phi_i \hat a^\dagger_i \hat a_i,
\end{equation}
where $\Phi_i$ are classical numbers, per s\'e (i.e. when neglecting the free hopping dynamics) does not generate entanglement, as it  simply adds a relative phase to the site-components of the statevector:
\begin{equation}
    \hat a^\dagger_i \rightarrow  e^{-i \Phi_i t/\hbar}\hat a^\dagger_i.
\end{equation}
Accordingly, $\hat U_{\text{int}}(t) = e^{-i \hat H_{\text{int}} t / \hbar}$ factorises with respect  to  ${\mathcal P}_2$ since $\hat H_{\text{int}}$ is  local with respect to that partition and, as such, it cannot generate spatial entanglement from a spatially separable state. In the present context, any spatial entanglement requires the kinetic or hopping term, which transports quantum amplitudes between the regions. The same is true for the specific case analyzed in Ref.~\cite{AzizHowl2025}. As shown in Sec.~\ref{Sec:TransitionAmplitudes} of the SM, entanglement is found in Ref.~\cite{AzizHowl2025} because the free evolution, which allows particles to diffuse from one object to the other, is not suppressed consistently at all places in the calculations. When this diffusion is suppressed throughout, the wavefunction is shown to remain factorized, like in the lattice-analogue considered here.

To summarize, the two apparently conflicting conclusions in the literature arise because the two entanglement analyses implicitly refer to different partitions of the Fock space. While $\ket{\Psi(t)}$ remains factorized with respect to the dynamically evolved mode partition, it is generally entangled with respect to the fixed spatial partition considered in~\cite{AzizHowl2025}. However, this spatial entanglement is generated by the free propagation of the particles across the chosen partition and is already present in the absence of the classical gravitational potential; the latter therefore does not act as the entangling mediator.  Clearly, its presence modifies the matrix coefficients, but this is what interactions generally do. 

The fact that, according to Eq.~(\!(9)\!) of~\cite{AzizHowl2025}, the 4$^{\text{th}}$ order exchange contribution to the transition amplitudes $\beta_{ij}(t)$ vanish when the gravitational potential is set to 0, naively suggesting that it is the classical gravitational potential that generates entanglement, is a simple consequence of Eq.~\eqref{eq:transition}, with $\hat U(t) = \hat{U}_0(t) \hat U_I(t)$. The authors~\cite{AzizHowl2025} compute $\hat U(t)$ by effectively neglecting $\hat U_0(t)$ while  retaining the free evolution of the matter fields in $\hat U_I(t)$, since it is an interaction picture-operator. When also the gravitational potential is set to 0, the dynamics becomes trivial with no time evolution and thus no entanglement generation. 

However, to assess whether it is the gravitational potential the true source of entanglement,  a good check is to remove the free evolution completely. (Note in fact that a  quantum gravitational interaction entangles systems also without their free kinetic term.)  As discussed here above and further detailed in Sec.~I  of SM, to neglect the free evolution {\it consistently} at all places, one should not only neglect $\hat U_0(t)$, but also the free evolution of the matter fields in $\hat U_I(t)$. Then, when only the classical gravitational interaction is retained, there is a nontrivial time evolution but no entanglement generation. Conversely, if the gravitational potential is removed and the free evolution retained, there is also entanglement generation---with respect to the chosen partition $\mathcal{P}_2$ of the Fock space. Thus it is the free evolution, not classical gravity, that generates entanglement. 

As a final comment, virtual particles, invoked by the authors to explain the entangling mechanism, do not play any role, since---as shown in Sec.~\ref{Sec:FirstQuantization} of SM---the very same calculation can be carried out in first quantitation, where virtual particles do not arise. The entangling role is indeed played by the free evolution. 

\section*{Acknowledgements}
The authors thank S. Donadi for discussions, and C. Karapoulitidis, V. Fragkos, K. Beyer, I. Pikovski for identifying an error in the previous version of the manuscript, which has been corrected. The authors acknowledge financial support from the University of Trieste, INFN and the EIC Pathfinder project QuCoM (GA No. 101046973).
\begin{widetext}
\begin{center}
\textbf{SUPPLEMENTARY MATERIAL}
\end{center}

\section{Fourth-order exchange contribution to the time evolution}\label{Sec:TransitionAmplitudes}

We show that the entanglement found in Ref.~\cite{AzizHowl2025, AzizHowl2025_supp} is due to diffusion of particles from one object to the other, and not the gravitational potential. In fact, if the free evolution is removed, but the potential retained, we show that the two objects do not become entangled. We also highlight some inconsistencies in the calculations performed in Ref.~\cite{AzizHowl2025, AzizHowl2025_supp}.

\subsection{Non-factorized form of the full wavefunction}\label{subsec:ExchangeTerms_QFT}
The factorized initial state considered in Ref.~\cite{AzizHowl2025_supp}, in second quantization, reads
\begin{align}\label{eq:InitialState_Fact}
\ket{\Psi(0)} &= \frac{1}{2}\sum_{m}\ket{N}_{1m}\sum_{k}\ket{N}_{2k}
=\frac{1}{2\sqrt{N!}}\sum_{m}(\hat{A}_{1m}^\dagger)^N
\frac{1}{\sqrt{N!}}\sum_{k}(\hat{A}_{2k}^\dagger)^N\ket{0},
\end{align}
where we have defined the operators
\begin{equation}
\hat{A}_{am}^\dagger :=  \frac{1}{c\sqrt{\hbar V}}\int d\mathbf{x}\theta_{am}(\mathbf{x})\hat{\phi}^{(+)\dagger}(0,\mathbf{x}), \qquad a\in \{1,2\}\,, m\in \{L,R\},
\end{equation}
with $\hat{\phi}^{(+)}$ being the positive frequency part of the complex Klein-Gordon field operator $\hat{\phi}$ such that
\begin{align}
    \hat{\phi}^{(+){\dagger}}(0,\mathbf{x}) = \frac{c\sqrt{\hbar}}{(2\pi)^3}\int \frac{d\mathbf{k}}{\sqrt{2\omega_k}}\hat{a}^{\dagger}_{\mathbf{k}}e^{-i\mathbf{k}\cdot \mathbf{x}}.
\end{align}
See, for instance, the discussion around Eqs.~(\!(7)\!) and~(\!(9)\!) in Ref.~\cite{AzizHowl2025_supp}. Here, we are using the notation where the equation numbers within the double parentheses $(\!(\cdot)\!)$ refer to the equation numbers in Ref. \cite{AzizHowl2025_supp}. The operator $\hat{A}^\dagger_{am}$  thus creates a particle localized within the sphere centered at the coordinate $\bm{X}_{am}$.

The full time evolution operator is $\hat U_0\hat U_I$. The wavefunction at a later time $t$, to fourth order, without making any further approximations, is determined by the following time evolution operator
\begin{equation}
\ket{\Psi^{[4]}(t)} = \hat{U}_0(t)\hat U_I^{(4)}(t)\frac{1}{2}\sum_{m}\ket{N}_{1m}\sum_{k}\ket{N}_{2k},
\end{equation}
where
\begin{equation}
\hat U_I^{(4)}(t)
=\frac{1}{4!}\left(-\frac{i}{\hbar}\right)^4
T\!\left[\prod_{r=1}^{4}\int_0^t d\tau_r\,
\hat H_I(\tau_r)\right],
\end{equation}
and the interaction Hamiltonian is given by Eq.~(\!(42)\!) in \cite{AzizHowl2025_supp} to be:
\begin{equation}\label{eq:IntHam}
\hat H_I
=
\frac{4}{c^2}\int d^3\mathbf{x}\,\Phi(\mathbf{x})
\left(
\hat\pi(x)\hat\pi^\dagger(x)
-
\frac{m^2c^2}{2\hbar^2}\hat\phi^\dagger(x)\hat\phi(x)
\right),
\end{equation}
where  $\hat\pi(x)$ is the conjugate momentum corresponding to $\hat\phi(x)$, and $\Phi(\mathbf{x})$ the gravitational potential, which is supposed to be {\it classical}. 

The interaction Hamiltonian contains a term proportional to $\int d\mathbf{x}\Phi(\mathbf{x})\hat \phi^{\dagger}(x) \phi(x)$, and another similar term involving the conjugate momentum. The action of the latter on the states can be computed analogously to the term proportional to $\hat{\phi}^{\dagger}(x)\hat\phi (x)$ as detailed after Eq.~$(\!(24)\!)$. We therefore focus on the  mass density term in $\hat{H}_I$, and include the contribution coming from the conjugate momentum term later.

We now discuss the so-called exchange terms, which in Ref.~\cite{AzizHowl2025_supp} are ascribed to entanglement generation.  These exchange terms, as specified by the contractions in Eq.~(\!(61)\!), involve destroying two particles, one from each object, and creating them again at later times (to preserve the particle number) at different locations. At fourth order, the interaction Hamiltonian has four creation and four annihilation operators. Two of these four annihilation operators act on the initial state (combinatorial factor $4\times3=12$), and the remaining two are contracted with two creation creation operators at locations different from the annihilation operators themselves (combinatorial factor $2\times1=2$). We thus have $4!$ terms of the type
\begin{align}\label{eq:Contractions}
\hat{\phi}^{\dagger}(x)\hat{\phi}^{\dagger}(y)\Phi(\mathbf{x})\contraction{}{\hat\phi}{(x)}{\hat\phi^\dagger}
\hat\phi(x)\hat\phi^\dagger(z)\Phi(\mathbf{z})\contraction{}{\hat\phi}{(w)}{|1\rangle}\hat\phi(z)|N\rangle_{1m}\Phi(\mathbf{y})\contraction{}{\hat\phi}{(y)}{\hat\phi^\dagger}
\hat\phi(y)\hat\phi^\dagger(w)\Phi(\mathbf{w})\contraction{}{\hat\phi}{(w)}{|1\rangle}\hat\phi(w)|N\rangle_{2k}.
\end{align}
Then, using the relation as in Eq.~$(\!(22)\!)$,
\begin{align}\label{eq:contraction_ket}
\contraction{}{\hat\phi}{(w)}{|1\rangle}\hat\phi(z)|N\rangle_{am} = \frac{c\hbar\sqrt{N}}{\sqrt{V}}\tilde{\theta}_{a m}(z)|N-1\rangle_{am},\qquad \tilde{\theta}_{am}(z):= \int d\mathbf{k}\frac{e^{ik\cdot z}}{(2\pi)^3 \sqrt{2\omega_k}}\theta_{am}(\mathbf{k}),
\end{align}
the expression in Eq.~$(\!(63)\!)$ for the propagator
\begin{align}
\contraction{}{\hat\phi}{(x)}{\hat\phi^\dagger}
\hat\phi(x)\hat\phi^\dagger(z):= \Delta_F(x-z) = c\hbar
\int \frac{d^4p}{(2\pi)^4}
\frac{i}{p^2+\gamma^2+i\epsilon}\,
e^{ip\cdot(x-z)},
\qquad
\gamma:=\frac{mc}{\hbar},     
\end{align}
and defining
\begin{align}\label{eq:CurlyA}
 \mathcal{A}^{\dagger}_{1m}(t)&:=\frac{1}{c^2\hbar^2}\left(\frac{2m^2}{\hbar^2}\right)^2\frac{\hbar c\sqrt{N}}{\sqrt{V}}\,
\int_t d^4xd^4z \hat{\phi}^{\dagger}( x)\Phi(\mathbf{x})\Phi(\mathbf z)\Delta_F(x-z)\tilde{\theta}_{1m}( z),\nonumber\\
\mathcal{A}^{\dagger}_{2k}(t)&:= \frac{1}{c^2\hbar^2}\left(\frac{2m^2}{\hbar^2}\right)^2\frac{\hbar c\sqrt{N}}{\sqrt{V}}\int_t d^4wd^4y\hat{\phi}^{\dagger}( y)\Phi(\mathbf{y})\Phi(\mathbf w)\Delta_F(y-w)\tilde{\theta}_{2k}(w),
\end{align}
where $\int_t d^4xd^4z:= \int_{0}^{ct} dx_0\int_{0}^{ct} dz_0\int d\mathbf{x}d\mathbf{z}$, we get
\begin{align}
 \hat{U}_{I;\text{ex}}^{(4)}(t)\ket{N}_{1m}\ket{N}_{2k} = \hat {\mathcal{A}}^{\dagger}_{1m}(t)\hat{\mathcal{A}}^{\dagger}_{2k}(t)\ket{N-1}_{1m}\ket{N-1}_{2k}.   
\end{align}
With the unitary time evolution taken into account, we get for the final state
\begin{align}\label{eq:FinalState_Fact}
 \ket{\Psi^{[4]}_{\text{ex}}(t)}= \hat{U}_0\hat{U}_{I;\text{ex}}^{(4)}(t)\frac{1}{2}\sum_{m}\ket{N}_{1m}\sum_{k}\ket{N}_{2k}.
\end{align}
Further, for computing the exchange terms, by definition, one chooses those terms where at a later time $t$ the operator $\hat{\mathcal{A}}^{\dagger}_{1m}$  creates one particle back only in one of the superposed states of object 2: $\ket{N-1}_{2k}$. Similarly, $\hat{\mathcal{A}}^{\dagger}_{2k}$ acts on one of the kets of object 1: $\ket{N-1}_{1m}$. This can also be seen from the contractions specified in Eq.~(\!(61)\!) in Ref.~\cite{AzizHowl2025_supp}. Thus, at a later time $t$, the state $\ket{\Psi^{[4]}_{\text{ex}}(t)}$  has $N$ particles again in each of the superposed branches, and reads
\begin{align}\label{eq:FinalState_Fact_N}
\ket{\Psi^{[4]}_{\text{ex}}(t)}= \frac{1}{2}\hat{U}_0\sum_{m,k}\hat {\mathcal{A}}^{\dagger}_{1m}(t)\ket{N-1}_{2k}\hat{\mathcal{A}}^{\dagger}_{2k}(t)\ket{N-1}_{1m}.  
\end{align}

We thus see that the exchange terms, for $N>1$, do not maintain the factorized form  of the statevector. That is, the sums over $m$ and $k$ cannot be factorized for $N>1$. However, this is not true for $N=1$. Indeed, in this case we have
\begin{align}\label{eq:FinalState_Fact_1}
\ket{\Psi^{[4]}_{\text{ex}}(t)}|_{N=1}= \frac{1}{2}\hat{U}_0\left(\sum_{m}\hat {\mathcal{A}}^{\dagger}_{1m}(t)\sum_{k}\hat{\mathcal{A}}^{\dagger}_{2k}\right)\ket{0}.  
\end{align}

Note that in Ref.~\cite{AzizHowl2025_supp}, after further ignoring $\hat{U}_0$ in Eq.~\eqref{eq:FinalState_Fact_N}, the non-factorized form of the exchange contribution is reported, but not the fact that $\ket{\Psi^{[4]}_{\text{ex}}(t)}|_{N=1}$ remains factorized. One would expect that the conclusion about the final state being entangled or not, must not change drastically in going from $N=1$ to $N>1$. This seems to suggest that the exchange terms, by themselves, may not be sufficient to describe and quantify the entangled structure of the wavefunction.  

Indeed, if one  considers the time evolution of the full state, instead of focusing only on the exchange terms, it can be seen from the discussion in Sec.~\ref{Sec:FirstQuantization} that for $N=1$, two classes of terms contribute to the evolution of the full wavefunction: the direct terms, which correspond to the left terms of the coefficients computed in Eqs.~\eqref{eq:aLL}-\eqref{eq:aRR}, and the exchange terms, which correspond to the right terms in Eqs.~\eqref{eq:aLL}-\eqref{eq:aRR}. For $N=1$, while both the direct and exchange terms separately preserve the factorized form of the wavefunction---c.f. Eq.~\eqref{eq:fact_dist} for factorization of direct terms and Eq.~\eqref{eq:factorized_ex_1} for factorization of exchange terms---the full wavefunction becomes a sum of two factorized pieces, which itself is not factorized. See also the discussion around Eq.~\eqref{eq:Direct+Exchange} in Sec.~\ref{sec:Lattice}. Thus, the full state, both for $N=1$ and $N>1$, remains non-factorized. This also implies that while the direct terms, by themselves, always preserve the factorized structure of the wavefunction, it is not true that they can be ignored in discussions related to entanglement, contrary to the discussion in Ref.~\cite{AzizHowl2025_supp} after Eq.~(\!(55)\!).

The relevant question now is to understand the origin of the entanglement. We show in the next subsection~\ref{subsec:scattering}, by explicitly computing the exchange coefficients as defined in Eq.~(\!(61)\!), that the entanglement found by the authors~\cite{AzizHowl2025_supp} is actually due to the transition of one particle from object 1 to object 2  and another particle from object 2 to object 1. This is in contrast with the physical situation that the authors~\cite{AzizHowl2025_supp} consider, where such processes are not allowed, and objects are treated like rigid bodies not leaking particles away. We quote the sentences below Eq.~(\!(38)\!) of Ref.~\cite{AzizHowl2025_supp}: ``\textit{...This is because it is not possible for one atom of one object to freely diffuse from one object to the other. That is, we assume that there are no “direct” interactions between the two matter systems...}"

Therefore, as will become clear shortly, the physical situation which the authors have in mind is inconsistent with what they compute and attribute to entanglement. We show further in the next subsection that if one truly considers processes where particles are not allowed to transition from one object to another, by suppressing the free evolution consistently at all places, the full wavefunction remains factorized, contrary to the conclusion of Ref.~\cite{AzizHowl2025_supp}.

\subsection{Transition of particles between the two objects}\label{subsec:scattering}
In this section we compute the transition amplitudes $\beta^{(4);\rm{ex}}_{ij;mk}(t):= {}_{1i}\bra{N} {}_{2j}\bra{N}\hat{U}_0\hat{U}_{I;\text{ex}}^{(4)}(t)\ket{N}_{1m}\ket{N}_{2k}$ using the same approximations as in Ref.~\cite{AzizHowl2025_supp}. We show that the exchange transition amplitudes attributed to entanglement in Ref.~\cite{AzizHowl2025_supp}, in the language of Feynman diagrams, correspond to the transition of identical particles from object 1 to object 2 and vice versa. 

Using Eq.~\eqref{eq:FinalState_Fact} and Eq.~\eqref{eq:CurlyA}, the exchange coefficients are given by
\begin{align}\label{eq:BetaEx}
\beta^{(4);\rm{ex}}_{ij;mk}(t)
&={}_{1i}\bra{N} {}_{2j}\bra{N}\hat{U}_0\hat {\mathcal{A}}^{\dagger}_{1m}(t)\hat{\mathcal{A}}^{\dagger}_{2k}(t)\ket{N-1}_{1m}\ket{N-1}_{2k} \nonumber\\
&=\frac{1}{c^4\hbar^4}\left(\frac{2m^2}{\hbar^2}\right)^4\frac{\hbar^2 c^2 N}{V}\,
\int_t d^4{x}d^{4}{y}d^4{w}d^4{z}\Phi(\mathbf x)\Phi(\mathbf y)\Phi(\mathbf w)\Phi(\mathbf z)\tilde{\theta}_{1m}(z)\tilde{\theta}_{2k}(w)\times\nonumber\\
&\times\Delta_F(x-z)\Delta_F(y-w)
_{1i}\!\langle N|\,{}_{2j}\!\langle N|\,\hat U_0(t)\hat{\phi}^{\dagger}(x)\hat{\phi}^{\dagger}(y)\,|N-1\rangle_{1m}|N-1\rangle_{2k}.
\end{align}
Next, Ref.~\cite{AzizHowl2025_supp} makes an approximation concerning the treatment of $\hat{U}_0$ in Eq.~\eqref{eq:BetaEx}. The approximation corresponds to treating ${}_{1i}\bra{N} {}_{2j}\bra{N}$ as an eigenstate of $\hat{U}_0$, such that 
\begin{align}
{}_{1i}\bra{N} {}_{2j}\bra{N}\hat{U}_0 \approx  {}_{1i}\bra{N} {}_{2j}\bra{N}e^{i2Mc^2/\hbar}.   
\end{align}
In the same spirit, in computing the contraction in Eq.~\eqref{eq:contraction_ket}, the spatial spread of the initial statevector due to the action of the free evolved operator $\hat{\phi}$, is also ignored, such that Eq.~\eqref{eq:contraction_ket} becomes (cf.~Eq.~(\!(23)\!))
\begin{align}\label{eq:contraction_ket_nofree}
\contraction{}{\hat\phi}{(w)}{|1\rangle}\hat\phi(z)|N\rangle_{am} \approx\frac{\hbar\sqrt{N}}{\sqrt{2mV}}e^{-imcz^0/\hbar}{\theta}_{a m}(\mathbf{z})|N-1\rangle_{am}\,.
\end{align}
To compute the exchange terms connecting objects 1 and 2,  we must contract the operators $\hat{\phi}^{\dagger}(x)$ and $\hat{\phi}^{\dagger}(y)$ with the appropriate bras. The chosen contractions are:
\begin{align}
\contraction{{}_{1i}\!}{\langle N|}{}{\phi^\dagger}
{}_{1 i}\!\langle N|\hat\phi^\dagger(y)
\approx
\frac{\hbar\sqrt{N}}{\sqrt{2m V}}\,
e^{+imcy^0/\hbar}\,
\theta_{1i}(\mathbf y)\,
{}_{1i}\!\langle N-1|,\qquad \contraction{{}_{2j}\!}{\langle N|}{}{\phi^\dagger}
{}_{2 j}\!\langle N|\hat\phi^\dagger(x)
\approx
\frac{\hbar\sqrt{N}}{\sqrt{2m V}}\,
e^{+imcx^0/\hbar}\,
\theta_{2j}(\mathbf x)\,
{}_{2j}\!\langle N-1|.
\label{eq:ext-bra}
\end{align}

With these approximations, to fourth order, the exchange coefficients are given by
\begin{align}
\beta^{(4);\rm{ex}}_{ij;mk}
&=
{}_{1i}\bra{N-1}{}_{2j}\braket{N-1|N-1}_{1m}\ket{N-1}_{2k}\times\frac{4m^6N^2}{\hbar^6c^2V^2}
\int_t d^4x\,d^4y\,d^4z\,d^4w\;
\Phi(\mathbf x)\Phi(\mathbf y)\Phi(\mathbf z)\Phi(\mathbf w)\times\nonumber\\
&\qquad
\times e^{i\gamma(x^0-z^0)}e^{i\gamma(y^0-w^0)}\theta_{2j}(\mathbf x)\theta_{1m}(\mathbf z)\theta_{2k}(\mathbf w)\theta_{1i}(\mathbf y)
\int\frac{d^4p}{(2\pi)^4}\frac{d^4q}{(2\pi)^4}\frac{i\,e^{ip\cdot(x-z)}}{p^2+\gamma^2+i\epsilon}\cdot \frac{i\,e^{iq\cdot(y-w)}}{q^2+\gamma^2+i\epsilon}.
\label{eq:phi-only-general}
\end{align}

The momentum-conjugate contractions are handled exactly as in~\cite{AzizHowl2025_supp}:
for propagators involving \(\hat\pi\), the time derivatives generate extra factors of
\(i\gamma\) or \(\gamma^2\). These modify only the overall coefficient of each exchanged line,
not the branch-label structure. Accordingly, the full exchange contribution has the same branch dependence as Eq.~\eqref{eq:phi-only-general}.  

Now, to see the physics behind the coefficients in Eq.~\eqref{eq:phi-only-general}, it is instructive to see again the chosen contractions all together. They are:

\begin{equation}\label{eq:ExchangeContractions}
\contraction{}{\hat\phi}{(z)}{|N\rangle}
\hat\phi(z)|N\rangle_{1 m},\quad \Phi(\mathbf{x})\contraction{}{\hat\phi}{(x)}{\hat\phi^\dagger}
\hat\phi(x)\hat\phi^\dagger(z)\Phi(\mathbf{z}),
\quad \contraction{{}_{2 j}\!}{\langle N|}{}{\phi^\dagger}
{}_{2 j}\!\langle N|\hat\phi^\dagger(x),\qquad \contraction{}{\hat\phi}{(x)}{|N\rangle}
\hat\phi(w)|N\rangle_{2 k},\quad 
\Phi(\mathbf{w})\contraction{}{\hat\phi}{(w)}{\hat\phi^\dagger}
\hat\phi(y)\hat\phi^\dagger(w)\Phi(\mathbf{y}),\quad \contraction{{}_{N i}\!}{\langle N|}{}{\phi^\dagger}
{}_{1i}\!\langle N|\hat\phi^\dagger(y).
\end{equation}

\begin{figure}[h]
\begin{tabular}{ll}
\includegraphics[scale=0.6]{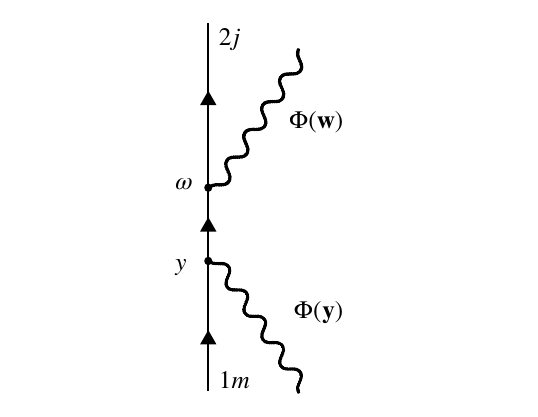}
&
\includegraphics[scale=0.6]{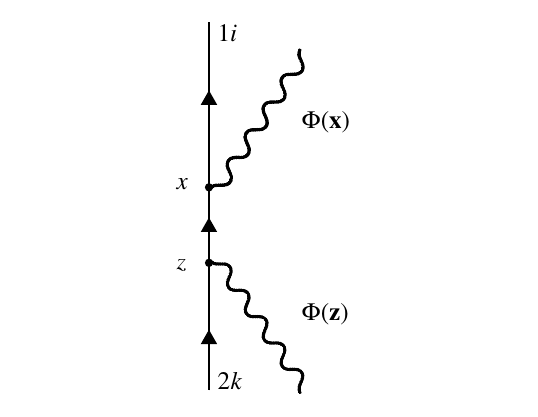}
\end{tabular}
\caption{Left: one particle (straight line) transitioning from object 1 at $\bm{X}_{1m}$ to object 2 at $\bm{X}_{2j}$, while interacting with the external classical gravitational potential (wiggly line). Right: one particle transitioning from object 2 at $\bm{X}_{2k}$ to object 1 at $\bm{X}_{1i}$ during the interaction with the external gravitational potential. This process is analogous to Compton scattering in quantum electrodynamics~\cite{GreinerChapter9}.}
\label{Fig:Feynman}
\end{figure}

In the language of Feynman diagrams (see Fig.~\ref{Fig:Feynman}), the first three contractions in Eq.~\eqref{eq:ExchangeContractions} correspond to the transition of a particle from object 1 centered at $\bm{X}_{1m}$ with $(m \in {L,R})$ to object 2 centered $\bm{X}_{2j}$ with $(j \in {L,R})$.  During the transition, the particle interacts with the external gravitational potential, and evolves  with the free propagator $\contraction{}{\hat\phi}{(x)}{\hat\phi^\dagger}
\hat\phi(x)\hat\phi^\dagger(z)$. The last three contractions in Eq.~\eqref{eq:ExchangeContractions} describe the same process where the particle transitions from object 2 centered at $\bm{X}_{2k}$ $(k \in {L,R})$ to object 1 centered $\bm{X}_{1i}$ $(i \in {L,R})$.

These transitions are not a prerogative of the QFT treatment; in Sec.~\ref{Sec:FirstQuantization}, we perform the same calculation within first quantization, re-obtaining the very same result.   Therefore,  these transition amplitudes cannot be understood as being intrinsic to virtual particle exchange, as suggested by the authors~\cite{AzizHowl2025}, since virtual particles do not arise in first quantization.
 
 We now evaluate the two exchanged propagators exactly as in Eqs.~(\!(64)\!)--(\!(69)\!).
Concerning $\Delta_{F}(x-z)$, using Eq.~(\!(64)\!), we have
\begin{equation}
\int \frac{d^3\mathbf p}{(2\pi)^3}
\frac{e^{i\mathbf p\cdot(\mathbf x-\mathbf z)}}{\mathbf p^2-(p^0)^2+\gamma^2}
=
f(p^0;\mathbf x-\mathbf z),
\end{equation}
with \(f\) given explicitly in Eq.~(\!(65)\!). Then the corresponding time integral is
\begin{align}
\mathcal K(\mathbf x,\mathbf z,t)
&:=
\int_0^{ct} dx^0 \int_0^{ct} dz^0\;
e^{i\gamma(x^0-z^0)}
\int \frac{dp^0}{2\pi}\,
f(p^0;\mathbf x-\mathbf z)\,
e^{-ip^0(x^0-z^0)}
\nonumber\\
&\approx
\frac{ct}{4\pi |\mathbf x-\mathbf z|}.
\label{eq:kernel-xz}
\end{align}
Similarly, for the second propagator we get
\begin{equation}
\mathcal K(\mathbf w,\mathbf y,t)\approx \frac{ct}{4\pi |\mathbf w-\mathbf y|}.
\label{eq:kernel-wy}
\end{equation}
Substituting Eqs.~\eqref{eq:kernel-xz} and \eqref{eq:kernel-wy} into
Eq.~\eqref{eq:phi-only-general}, and then restoring the contributions from the \(\hat\pi\)-insertions
exactly as in the paragraph immediately before Eq.~(\!(70)\!), we obtain
\begin{align}
\beta^{(4);\rm{ex}}_{ij;mk}&\approx
\frac{N^2m^6 t^2}{4\pi^2\hbar^6 V^2}
{}_{1i}\bra{N-1}{}_{2j}\braket{N-1|N-1}_{1m}\ket{N-1}_{2k}\int d^3\mathbf x\,d^3\mathbf z\,d^3\mathbf y\,d^3\mathbf w\;
\frac{\Phi(\mathbf x)\Phi(\mathbf z)\Phi(\mathbf y)\Phi(\mathbf w)}
{|\mathbf x-\mathbf z|\,|\mathbf w-\mathbf y|}
\nonumber\\
&\qquad\times\theta_{2j}(\mathbf x)\theta_{1m}(\mathbf z)\theta_{2k}(\mathbf w)\theta_{1i}(\mathbf y)
\times i^2,
\label{eq:M4-preI}
\end{align}
which factorizes for $N=1$ as
\begin{align}
\beta^{(4);\rm{ex}}_{ij;mk}&\approx
\frac{m^6 t^2}{4\pi^2\hbar^6V^2}
\left(
i\int d\mathbf x\,d\mathbf z\;
\frac{\Phi(\mathbf x)\Phi(\mathbf z)\,\theta_{2j}(\mathbf x)\theta_{1m}(\mathbf z)}
{|\mathbf x-\mathbf z|}
\right)
\left(
i\int d\mathbf w\,d\mathbf y\;
\frac{\Phi(\mathbf w)\Phi(\mathbf y)\,\theta_{2k}(\mathbf w)\theta_{1i}(\mathbf y)}
{|\mathbf w-\mathbf y|}
\right),
\label{eq:M4-factorized}
\end{align}
while for $N>1$ we get
\begin{align}
\beta^{(4);\rm{ex}}_{ij;mk}&\approx
\frac{N^2m^6 t^2}{4\pi^2\hbar^6V^2}
\left(
i\int d\mathbf x\,d\mathbf z\;
\frac{\Phi(\mathbf x)\Phi(\mathbf z)\,\theta_{2j}(\mathbf x)\theta_{1m}(\mathbf z)}
{|\mathbf x-\mathbf z|}
\right)\delta_{im}\delta_{jk}
\left(
i\int d\mathbf w\,d\mathbf y\;
\frac{\Phi(\mathbf w)\Phi(\mathbf y)\,\theta_{2k}(\mathbf w)\theta_{1i}(\mathbf y)}
{|\mathbf w-\mathbf y|}
\right).
\label{eq:M4-factorized_NGreater1}
\end{align}
In Eqs.~\eqref{eq:M4-factorized} and~\eqref{eq:M4-factorized_NGreater1} we have used the relation
\begin{align}\label{eq:orthonormality}
{}_{1i}\bra{N-1}{}_{2j}\braket{N-1|N-1}_{1m}\ket{N-1}_{2k} &= \delta_{im}\delta_{jk}\qquad \text{for } N>1,\nonumber\\
{}_{1i}\bra{N-1}{}_{2j}\braket{N-1|N-1}_{1m}\ket{N-1}_{2k} &= \braket{0|0} = 1\quad \text{for } N=1.
\end{align}
Renaming dummy integration variables, and using the symmetry of the denominator,
we define
\begin{equation}
V_{ij}
:=
i\int d^3\mathbf x\,d^3\mathbf y\;
\frac{\Phi(\mathbf x)\Phi(\mathbf y)\,\theta_{1i}(\mathbf x)\theta_{2j}(\mathbf y)}
{|\mathbf x-\mathbf y|},
\end{equation}
so that Eqs.~\eqref{eq:M4-factorized} and~\eqref{eq:M4-factorized_NGreater1} become, respectively
\begin{equation}
\beta^{(4);\rm{ex}}_{ij;mk}\big|_{N=1}\approx
\frac{m^6 t^2 }{4\pi^2\hbar^6V^2}\,
V_{mj}V_{ik},\qquad \beta^{(4);\rm{ex}}_{ij;mk}\big|_{N>1}\approx
\frac{N^2m^6 t^2 }{4\pi^2\hbar^6V^2}\,
V^2_{ij}.\label{eq:M4-final_1}
\end{equation}
On the projected space, the  wavefunction is written as
\begin{align}
    \ket{\Psi^{[4]}_{\text{ex}}(t)} = \sum_{ijm k}\frac{1}{2}\beta^{(4);\rm{ex}}_{ij;mk}(t)\ket{N}_{1i}\ket{N}_{2j}.
\end{align}
Then, for $N=1$, using Eq.~\eqref{eq:M4-final_1}, we see that the exchange piece of the wavefunction stays factorized
\begin{align}\label{eq:factorized_ex_1}
\ket{\Psi^{[4]}_{\text{ex}}(t)} \big|_{N=1}= \frac{1}{2}\frac{m^6 t^2 }{4\pi^2\hbar^6V^2}\sum_{i}\left(\sum_{k}V_{ik}\right)\ket{1}_{1i}\sum_{j}\left(\sum_{m}V_{mj}\right)\ket{1}_{2j},  
\end{align}
while for $N>1$ it becomes non-factorized
\begin{align}
\ket{\Psi^{[4]}_{\text{ex}}(t)} \big|_{N>1}= \frac{1}{2}\frac{N^2m^6 t^2 }{4\pi^2\hbar^6V^2}\sum_{ij}V^2_{ij}\ket{N}_{1i}\ket{N}_{2j}. 
\end{align}
We now discuss these results.

As we described earlier after Eq.~\eqref{eq:FinalState_Fact_1}, concerning the question of whether or not the statevector remains factorized as intended in Ref.~\cite{AzizHowl2025}, it is not relevant whether or not the exchange terms themselves have a factorized form, but rather, whether or not they are zero, since when they are zero, we are only left with the direct terms, which maintain the factorized form of the wavefunction. If they are not zero, the state in general is entangled.

In other words, it is the transition of particles from one object to the other, as captured by the exchange terms, which leads to the non-factorized form of the wavefunction found in Ref.~\cite{AzizHowl2025}. If such processes are discarded, say due to experimental scenarios of interest where the two objects are treated as rigid bodies not letting particles out due to a binding potential, the full wavefunction remains factorized.
This can be seen by considering the situation where the free evolution is neglected entirely, $\hat{H}_{\text{free}}=0$, such that only the interaction Hamiltonian $\hat{H}_{I}$  in Eq.~\eqref{eq:IntHam} determines the evolution of the full statevector (not just the exchange part).  We first consider the action of  $\hat{H}_{I}$ on the initial state in the Schr\"{o}dinger picture. We have 
\begin{align}\label{eq:NoFreeEvolution}
-\frac{i}{\hbar}\hat{H}_{I}\ket{\Psi(0)} =  i\frac{2m^2}{\hbar^3}\frac{1}{2}\sum_{n,k}\int d\mathbf{x}\,\Phi(\mathbf{x})
\hat\phi^\dagger(0,\mathbf{x})\hat\phi(0,\mathbf{x})\ket{N}_{1n}\ket{N}_{2k}.
\end{align}
Then, using Eq.~(\!(23)\!) of Ref.~\cite{AzizHowl2025_supp}, or Eq.~\eqref{eq:contraction_ket_nofree}, which is valid when the free evolution of $\hat{\phi}(x)$ is neglected, we get
\begin{align}\label{eq:NoFreeEvolution}
i\frac{2m^2}{\hbar^3}\int d\mathbf{x}\,\Phi(\mathbf{x})
\hat\phi^\dagger(0,\mathbf{x})\hat\phi(0,\mathbf{x})\ket{N}_{1n}\ket{N}_{2k} = i\frac{N m}{\hbar V}\int d\mathbf{x}\,\Phi(\mathbf{x})
\left(\theta_{1n}(\mathbf{x})+\theta_{2k}(\mathbf{x})\right)\ket{N}_{1n}\ket{N}_{2k},
\end{align}
such that
\begin{align}\label{eq:Factorization_Nofree}
\ket{\Psi(t)} &= \exp{\left(-i\hat{H}_{I}t/\hbar\right)}\ket{\Psi(0)}\nonumber\\ 
&=  \frac{1}{2}\sum_{n} \exp{\left(i\frac{ N m t}{\hbar V}\int d\mathbf{x}\,\Phi(\mathbf{x})
\theta_{1n}(\mathbf{x})\right)}\ket{N}_{1n}\sum_{k}\exp{\left(i\frac{N m t}{\hbar V}\int d\mathbf{x}\,\Phi(\mathbf{x})
\theta_{2k}(\mathbf{x})\right)}\ket{N}_{2k}.
\end{align}
Eq.~\eqref{eq:Factorization_Nofree} shows that if the free evolution is neglected consistently at all places, the full wavefunction at a later time $t$ remains factorized.

Instead, Ref.~\cite{AzizHowl2025_supp} finds a non-factorized form of the wavefunction. This is because diffusion of particles due to free evolution is neglected only at certain places (cf.~Eqs.~(\!(14)\!) and~(\!(24)\!) in Ref.~\cite{AzizHowl2025_supp}), and retained at other places, for instance, in computing the propagator $\Delta_{F}(x-y)$ (cf.~Eq.~(\!(63)\!) in Ref.~\cite{AzizHowl2025_supp}).

If the goal is to assess entanglement as due to the presence of the interaction, and not due to particles propagating from one object to the other—in line with the physical situation that the authors~\cite{AzizHowl2025_supp} are interested in, as per the discussion following Eq.~(\!(38)\!) in Ref.~\cite{AzizHowl2025_supp}—the free evolution must be ignored consistently at all places, including the computation of the propagator $\Delta_{F}(x-y)$. 

The conclusion is that the diffusion of particles from one object to the other is responsible for the entanglement found in Ref.~\cite{AzizHowl2025_supp}, not the classical gravitational potential. 

Another way to see this is to consider objects 1 and 2  made of distinguishable particles, so that the diffusion from one object to the other can be easily tracked and removed. Similarly to the previous analysis, one considers the projection of the final wavefunction over the states $\ket{N_a}_{1i}\ket{N_b}_{2j}$, where $i,j\in \{L,R\}$ and $a$ and $b$ denote the two types of particles. In doing so, one  automatically neglects states such as $\ket{{(N-1)}_a, 1_b}_{1i}\ket{{(N-1)}_b,1_a}_{2j}$, which correspond to particle diffusion between the two objects; note that these states are implicitly retained when the particles are identical ($a=b$). The analysis  in Sec.~\ref{subsec:Distinguishable} shows that for  distinguishable particles the exchange coefficients are zero (cf.~Eq.~\eqref{eq:ExchangeZero}), and the full wavefunction remains factorized.

As a further note, the authors  attribute the exchange transition amplitudes to virtual particle exchange between the two objects. However, $\beta^{(4);\rm{ex}}_{ij;mk}$ should not be attributed to a process where particles never move from one object to the other, but only interact through virtual particle exchange. As stated before, they should be attributed to particle diffusion; this is most straightforwardly seen from the discussion of Sec.~\ref{subsec:Identical} where the same exchange coefficients are obtained, but this time staying within first quantization. Since virtual particles do not arise within the framework of the standard non-relativistic Schr\"{o}dinger equation, $\beta^{(4);\rm{ex}}_{ij;mk}$ cannot be attributed to a process intrinsic to virtual  particle exchange. 

The entanglement found in Ref.~\cite{AzizHowl2025} arises because of the free evolution, which allows particles to transition form one object to the other. The role of the gravitational potential is only to modify this transition amplitude.

\section{Distance dependence of the transition amplitudes $\beta^{(4);\rm{ex}}_{ij}$}\label{Sec:DistanceDependence}
Evaluating the exchange terms \eqref{eq:M4-factorized} requires computing the integral
\begin{align}
    \mathcal{I} = \int d\mathbf x\,d\mathbf z\;
\frac{\Phi(\mathbf{z})\Phi(\mathbf{x})\theta_{1i}(\mathbf{z})\theta_{2k}(\mathbf{x})}
{|\mathbf x-\mathbf z|}.
\end{align}
We compute $\mathcal{I}$ in the regime where the inter-branch distances $d_{ik}$, and the separation between the objects $\Delta x$, are much bigger than the radius of the objects $R$ such that $d_{ik}\gg R$ and $\Delta x\gg R$. When this is the case, the denominator $|{\mathbf{x}-\mathbf{z}}|$ inside the integral can be replaced with $d_{ik}$ and taken outside. This is because $\theta_{1i}(\mathbf{z})$ and $\theta_{2k}(\mathbf{x})$ restrict the coordinates to stay inside the two objects centered at $\bm{X}_{1i}$ and $\bm{X}_{2k}$ respectively. Therefore, since we have in mind the geometries where $|\bm{X}_{1i}-\bm{X}_{2k}| = d_{ik}\gg R$, the distance $|{\mathbf{x}-\mathbf{z}}|$ in $\mathcal{I}$ can be approximated by $|{\mathbf{x}-\mathbf{z}}|\approx d_{ik}$. Furthermore, the semiclassical potentials $\Phi(\mathbf{z})$ and $\Phi(\mathbf{x})$, as specified in Eq.~(\!(51)\!) of Ref.~\cite{AzizHowl2025_supp}, can be replaced with the potentials $\Phi_{1i}(\mathbf{z})$ and $\Phi_{2k}(\mathbf{x})$ inside the integral, where
\begin{align}
\Phi_{1i}(\mathbf{z})&:= -\frac{Gm}{2}\left(\frac{3}{2R}-\frac{|\mathbf{z}-\bm{X}_{1i}|^2}{2R^3} \right),\nonumber\\
\Phi_{2k}(\mathbf{x})&:=-\frac{Gm}{2}\left(\frac{3}{2R}-\frac{|\mathbf{x}-\bm{X}_{2k}|^2}{2R^3} \right).
\end{align}
That is, at locations $\mathbf{z}$ and $\mathbf{x}$ in $\mathcal{I}$, the predominant contribution to the semiclassical gravitational potential considered in Eq.~(\!(51)\!) of Ref.~\cite{AzizHowl2025_supp}, comes only from the spheres centered at $\bm{X}_{1i}$ and $\bm{X}_{2k}$ respectively. Making use of these considerations, in spherical polar coordinates, we get
\begin{align}
\mathcal{I} = \frac{4\pi^2G^2m^2}{d_{ik}}\left[\int_{0}^{R} dr r^2\left(\frac{3}{2R}-\frac{r^2}{2R^3} \right)\right]^2 = \frac{16\pi^2G^2m^2R^4}{25 d_{ik}}.
\end{align}
Plugging this back in Eq.~\eqref{eq:M4-factorized} we get
\begin{align}
\beta^{(4);\rm{ex}}_{ij;mk} = \frac{9 m^6 t^2}{4\pi^2\hbar^6 (4\pi)^2 R^6}\times \frac{i16\pi^2G^2m^2R^4}{25 d_{ik}}\times \frac{i16\pi^2G^2m^2R^4}{25 d_{mj}}, 
\end{align}
which yields the compact expression
\begin{align}\label{eq:BranchDistance}
\beta^{(4);\rm{ex}}_{ij;mk}=\left(\frac{6}{25}\frac{iG^2m^5Rt}{\hbar^3}\right)^2\frac{1}{d_{ik}d_{mj}}.
\end{align}
It is clear that for the diagonal terms where $m=i$ and $k=j$, we get the result obtained in Eq.~$(\!(84)\!)$ of Ref.~\cite{AzizHowl2025_supp} for $N=1$. The case $N>1$ can be computed similarly.

\section{Non perturbative solution for  fixed number of particles}\label{sec:NoEntanglement}
In this section we show that an initially factorized state remains factorized, as per the partition $\mathcal{P}_1$ discussed in the main text, when particles described by a (complex) Klein-Gordon field interact only with an external classical potential, and no-pair creation is allowed.

As before, $\hat\phi(x)$ is the complex Klein--Gordon field and  $\Phi(\mathbf{x})$  the prescribed external classical potential.  The free one-particle energy is
$E_p:=\sqrt{|\mathbf p|^2+m^2}$, and for simplicity we have set $\hbar=c=1$ throughout this section. In the interaction picture, the free field $\hat{\phi}(x)$ and its conjugate momentum are expanded as
\begin{align}
\hat\phi_I(t,\mathbf x)
&=
\int \frac{d^3p}{(2\pi)^{3/2}}\frac{1}{\sqrt{2E_p}}
\left(
\hat a(\mathbf p)e^{-iE_pt+i\mathbf p\cdot\mathbf x}
+
\hat b^\dagger(\mathbf p)e^{+iE_pt-i\mathbf p\cdot\mathbf x}
\right),\\
\hat\pi_I(t,\mathbf x)
&=\partial_t\hat\phi_I^\dagger(t,\mathbf x) =
i\int \frac{d^3p}{(2\pi)^{3/2}}\sqrt{\frac{E_p}{2}}
\left(
\hat a^\dagger(\mathbf p)e^{+iE_pt-i\mathbf p\cdot\mathbf x}
-
\hat b(\mathbf p)e^{-iE_pt+i\mathbf p\cdot\mathbf x}
\right).
\end{align}
In terms of the freely evolving fields, the interaction Hamiltonian considered in~\cite{AzizHowl2025_supp} is given by
\begin{equation}
\hat H^{\mathrm{int}}_{I}(t)
=4\int d^3\mathbf x\,\Phi(\mathbf x)
\left[
\hat\pi_I(t,\mathbf x)\hat\pi_I^\dagger(t,\mathbf x)
-\frac{m^2}{2}\hat\phi_I^\dagger(t,\mathbf x)\hat\phi_I(t,\mathbf x)
\right].
\end{equation}
In the Fourier space where
\begin{equation}
\widetilde\Phi(\mathbf k):=\int d^3\mathbf x\,e^{-i\mathbf k\cdot\mathbf x}\Phi(\mathbf x)
\end{equation}
denotes the spatial Fourier transform of the external potential, and using the mode expansions of $\hat{\phi}_I$ and $\hat{\pi}_I$,  $\hat H^{\mathrm{int}}_I(t)$ is given by
\begin{equation}
\label{eq:Hint-general}
\hat H^{\mathrm{int}}_{I}(t)
=
E(t)
+\hat a^\dagger \mathcal{A}_t\hat a
+\hat b^\dagger \mathcal{B}_t\hat b
+\hat a^\dagger \mathcal{C}_t\hat b^\dagger
+\hat b\,\mathcal{C}_t^\dagger\hat a.
\end{equation}
Here $E(t)$ is a c-number coming from normal ordering, and
\begin{align}
\hat a^\dagger \mathcal{A}_t\hat a
&=
\iint d^3\mathbf p\,d^3\mathbf q\,
\mathcal{A}_t(\mathbf p,\mathbf q)\,
\hat a^\dagger(\mathbf p)\hat a(\mathbf q),\\
\hat b^\dagger \mathcal{B}_t\hat b
&=
\iint d^3\mathbf p\,d^3\mathbf q\,
\mathcal{B}_t(\mathbf p,\mathbf q)\,
\hat b^\dagger(\mathbf p)\hat b(\mathbf q),\\
\hat a^\dagger \mathcal{C}_t\hat b^\dagger
&=
\iint d^3\mathbf p\,d^3\mathbf q\,
\mathcal{C}_t(\mathbf p,\mathbf q)\,
\hat a^\dagger(\mathbf p)\hat b^\dagger(\mathbf q),
\end{align}
where the functions $A_{t}, \mathcal{B}_t$ and $\mathcal{C}_t$ are given by
\begin{align}
\label{eq:A-kernel}
\mathcal{A}_t(\mathbf p,\mathbf q)
&=
\frac{\widetilde\Phi(\mathbf p-\mathbf q)}{(2\pi)^3}\,
\frac{2E_pE_q-m^2}{\sqrt{E_pE_q}}\,
e^{\,i(E_p-E_q)t},\\
\label{eq:B-kernel}
\mathcal{B}_t(\mathbf p,\mathbf q)
&=
\frac{\widetilde\Phi(\mathbf q-\mathbf p)}{(2\pi)^3}\,
\frac{2E_pE_q-m^2}{\sqrt{E_pE_q}}\,
e^{-\,i(E_p-E_q)t},\\
\label{eq:C-kernel}
\mathcal{C}_t(\mathbf p,\mathbf q)
&=
-\frac{\widetilde\Phi(\mathbf p+\mathbf q)}{(2\pi)^3}\,
\frac{2E_pE_q+m^2}{\sqrt{E_pE_q}}\,
e^{\,i(E_p+E_q)t}.
\end{align}
Since $\Phi$ is real, we have $\mathcal{B}_t=\mathcal{A}_t^\dagger$. There is no term of the form $\hat a^\dagger \mathcal{D}_t\hat b+\hat b^\dagger \mathcal{D}_t^\dagger\hat a$, because $\hat H^{\mathrm{int}}_I(t)$ in Eq.~\eqref{eq:Hint-general} commutes with the charge operator
\begin{equation}
\hat Q=\int d^3\mathbf p\,
\bigl(\hat a^\dagger(\mathbf p)\hat a(\mathbf p)-\hat b^\dagger(\mathbf p)\hat b(\mathbf p)\bigr).
\end{equation}
The terms $\hat a^\dagger \mathcal{C}_t\hat b^\dagger$ and $\hat b\,\mathcal{C}_t^\dagger\hat a$
create or annihilate a particle--antiparticle pair and therefore do not conserve the total number operator
\begin{equation}
\hat N=\int d^3\mathbf p\,
\bigl(\hat a^\dagger(\mathbf p)\hat a(\mathbf p)+\hat b^\dagger(\mathbf p)\hat b(\mathbf p)\bigr).
\end{equation}
Indeed, $[\hat N,\hat a^\dagger \mathcal{C}_t\hat b^\dagger]=2\hat a^\dagger \mathcal{C}_t\hat b^\dagger$ and
$ [\hat N,\hat b\,\mathcal{C}_t^\dagger\hat a]=-2\hat b\,\mathcal{C}_t^\dagger\hat a$.
As in~\cite{AzizHowl2025_supp}, we assume that the total number of particles remains constant (no pair-creation), therefore we neglect these terms and replace Eq.~\eqref{eq:Hint-general} by
\begin{equation}
\label{eq:Hint-np}
\hat H^{\text{int}}_I(t)
=
E(t)+\hat a^\dagger \mathcal{A}_t\hat a+\hat b^\dagger \mathcal{B}_t\hat b.
\end{equation}
Now, the particle and anti-particle sectors remain separate. Therefore, without loss of generality, we focus only on the former. We consider the one-particle Hilbert space $\mathcal H_1=L^2(\mathbb R^3,d^3\mathbf p)$. For $f\in\mathcal H_1$, we define $\hat a^{\dagger}(f)$ to be
\begin{equation}
\hat a^\dagger(f)=\int d^3\mathbf p\,f(\mathbf p)\hat a^\dagger(\mathbf p).
\end{equation}
For the states considered in~\cite{AzizHowl2025_supp}, $f(\mathbf{p})$ corresponds to the Fourier transform of the spatial distributions $\theta_{ai}/\sqrt{V}$. 
The Hamiltonian in Eq.~\eqref{eq:Hint-np} for the particle sector is the second quantization of the one-particle operator $\hat H_{1,I}^{\mathrm{int}}(t)$ defined as $\bra{\bf p}\hat H_{1,I}^{\mathrm{int}}(t)\ket{\bf q} =
\mathcal{A}_t({\bf p}, {\bf q})$: 
\begin{equation}\label{eq:dGamma-H1I}
\hat{H}^{\mathrm{int}}_I(t)=
 \hat a^{\dagger} \hat H_{1,I}^{\mathrm{int}} \hat a+E(t).
\end{equation}

To consider the full time evolution, we move to the Scr\"odinger picture, and add the free Hamiltonian $\hat{H}_0$
\begin{equation}
\hat H_0=\int d^3\mathbf p\,E_p\,
\hat a^\dagger(\mathbf p)\hat a(\mathbf p).
\end{equation}
Then, in the Schr\"odinger picture, the full (no-pair) Hamiltonian reads
\begin{equation}
\hat H(t)
=
\hat a^{\dagger} \hat H_{1} \hat a +E(t),
\end{equation}
with $\hat H_1$ defined as $\bra{\bf p} \hat H_1 \ket{\bf q} = \mathcal{A}_{t=0}({\bf p}, {\bf q}) + E_{\bf p} \delta^{(3)}({\bf p} - {\bf q})$.
Using the canonical commutation relations, one finds
\begin{align}
[\hat a^\dagger \mathcal{A}_t\hat a,\hat a^\dagger(f)]=\hat a^\dagger(\mathcal{A}_tf).
\end{align}
Here, as before, we use the compact notation 
\begin{align}
\mathcal{A}_tf:=\int d^3\mathbf{q}\mathcal{A}_t(\mathbf{p},\mathbf{q})f(\mathbf{q}).
\end{align}
Hence
\begin{equation}
\label{eq:comm-H-A}
[\hat H(t),\hat a^\dagger(f)]
=
\hat a^\dagger(\hat H_1 f)
\end{equation}
Let $\hat U(t,s)$ be the propagator generated by $\hat H (t)$,
\begin{equation}
i\partial_t\hat U(t,s)=\hat H(t)\hat U(t,s),
\qquad
\hat U(s,s)=1,
\end{equation}
and let $\hat U_1(t,s)$ be the one-particle propagator generated by $\hat H_1$,
\begin{equation}
i\partial_t\hat U_1(t,s)=\hat H_1\hat U_1(t,s),
\qquad
\hat U_1(s,s)=1.
\end{equation}
Let us define
\begin{equation}
\hat Y_f(t):=\hat U(t,s)^\dagger\,\hat a^\dagger(\hat U_1(t,s)f)\,\hat U(t,s).
\end{equation}
Using Eq.~\eqref{eq:comm-H-A}, one computes
\begin{align}
\partial_t\hat Y_f(t)
&=
i\hat U^\dagger \hat H(t)\hat a^\dagger(\hat U_1f)\hat U
+\hat U^\dagger \hat a^\dagger(\partial_t\hat U_1f)\hat U
-i\hat U^\dagger \hat a^\dagger(\hat U_1f)\hat H(t)\hat U
\notag\\
&=
i\hat U^\dagger [\hat H(t),\hat a^\dagger(\hat U_1f)]\hat U
-i\hat U^\dagger \hat a^\dagger(\hat H_1(t)\hat U_1f)\hat U
\notag\\
&=0.
\end{align}
Since $\hat Y_f(s)=\hat a^\dagger(f)$, it follows that
\begin{equation}
\label{eq:Heis-evolution}
\hat U(t,s)\hat a^\dagger(f)\hat U(t,s)^\dagger
=
\hat a^\dagger(\hat U_1(t,s)f).
\end{equation}
We highlight that by retaining the free evolution as per the right-hand-side of Eq.~\eqref{eq:Heis-evolution}, we go beyond the regime where the spread of the wavefunction due to the free evolution is suppressed. Moreover, since $\hat H(t)|0\rangle=E(t)|0\rangle$, the vacuum evolves only by a phase:
\begin{equation}
\label{eq:vac-phase}
\hat U(t,s)|0\rangle
=
e^{-i\theta(t,s)}|0\rangle,
\qquad
\theta(t,s)=\int_s^t E(\tau)\,d\tau.
\end{equation}
Consider a factorized bosonic Hartree state at the initial time $s$ 
\begin{equation}
\ket{\Psi(s)}=\frac{1}{\sqrt{(2N)!}} [\hat a^\dagger(f)]^{2N}|0\rangle.
\end{equation}
Then, using \eqref{eq:Heis-evolution} and \eqref{eq:vac-phase}, we obtain
\begin{align}
\ket{\Psi(t)}
=
\hat U(t,s)\ket{\Psi(s)}
& =
\frac{1}{\sqrt{(2N)!}}\,
\hat U(t,s) [\hat a^\dagger(f)]^{2N}\hat U(t,s)^\dagger\,
\hat U(t,s)|0\rangle
\nonumber \\
& =
e^{-i\theta(t,s)}
\frac{1}{\sqrt{(2N)!}}\,
[\hat a^\dagger\!\bigl(\hat U_1(t,s)f\bigr)]^{2N}|0\rangle.
\end{align}
Therefore, an initially factorized bosonic state remains factorized under the no-pair dynamics. More generally, if
\begin{equation}\label{eq:InStateQ2}
\ket{\Psi(s)}=\hat a^\dagger(f_1)\cdots \hat a^\dagger(f_{2N})|0\rangle,
\end{equation}
then
\begin{equation}
\ket{\Psi(t)}
=
e^{-i\theta(t,s)}
\hat a^\dagger\!\bigl(\hat U_1(t,s)f_1\bigr)\cdots
\hat a^\dagger\!\bigl(\hat U_1(t,s)f_{2N}\bigr)|0\rangle.
\end{equation}
Therefore, any symmetrized tensor product remains a symmetrized tensor product.

\section{The transition amplitudes in first quantization}\label{Sec:FirstQuantization}
In this section we show that the same expressions for the transition amplitudes $\beta_{ij}$ computed in Ref.~\cite{AzizHowl2025_supp} within second quantization, can also be derived  for {\it non-relativistic particles in first quantization}.
 
\subsection{Identical particles}\label{subsec:Identical}
Since the dynamics is non-relativistic and particle creation is not relevant, instead of the $N$-particle state considered in~\cite{AzizHowl2025}, we take a two-particle wavefunction where the first particle is in a superposition of the two different locations represented by $\ket{1L}$ and $\ket{1R}$ and also the second particle is in a superposition of the other two locations represented by $\ket{2L}$ and right $\ket{2R}$. We first consider the case of identical particles.
For two particles, the initially factorized state of \cite{AzizHowl2025_supp} written in second quantization (Q2) reads
\begin{align}
\ket{\Psi(t_i)} \overset{\mathrm{Q2}}{=} \frac{1}{2}\left(\ket{1}_{1L}+\ket{1}_{1R} \right)\otimes\left(\ket{1}_{2L}+\ket{1}_{2R}\right)\,.
\end{align}
Instead, in first quantization (Q1), it is given by
\begin{align}\label{eq:InitialQ1}
\ket{\Psi(t_i)} \overset{\mathrm{Q1}}{=} \frac{1}{2\sqrt{2}}\left[\left(\ket{1L}+\ket{1R}\right)\otimes\left(\ket{2L}+\ket{2R}\right)+\left(\ket{2L}+\ket{2R}\right)\otimes\left(\ket{1L}+\ket{1R}\right)\right]\,.
\end{align}
In order to model the wavepackets of the objects as in \cite{AzizHowl2025_supp}, the position states above represent non-overlapping smeared states:
\begin{align}\label{eq:wavepackets}
    \ket{1L}&= \frac{1}{\sqrt{V}}\int d^3\mathbf{x}\theta_{1L}(\mathbf{x})\ket{\mathbf{x}},\quad \ket{1R}= \frac{1}{\sqrt{V}}\int d^3\mathbf{x}\theta_{1R}(\mathbf{x})\ket{\mathbf{x}},\nonumber\\
    \ket{2L}&= \frac{1}{\sqrt{V}}\int d^3\mathbf{x}\theta_{2L}(\mathbf{x})\ket{\mathbf{x}},\quad 
    \ket{2R}= \frac{1}{\sqrt{V}}\int d^3\mathbf{x}\theta_{2R}(\mathbf{x})\ket{\mathbf{x}}.
    \end{align}
Next, in order to describe the dynamics considered in \cite{AzizHowl2025_supp}, but this time in first quantization, we take for the two-particle Hamiltonian
\begin{align}\label{eq:HamiltonianFQ}
\hat{H}=\frac{\hat{p}^2}{2m}\otimes\hat{I}+ \hat{I}\otimes \frac{\hat{p}^2}{2m} + \int d^3\mathbf{x}\left(m\Phi(\mathbf{x})\ket{\mathbf{x}}\bra{\mathbf{x}}\otimes\hat{I}+\hat{I}\otimes m\Phi(\mathbf{x})\ket{\mathbf{x}}\bra{\mathbf{x}}\right)\,.
\end{align}
When $\Phi(\mathbf{x})$ is treated classically, the unitary time evolution operator factorizes as 
\begin{align}\label{eq:Facto}
\hat{U}(t) = \hat{U}_s(t)\otimes \hat{U}_s(t),
\end{align}
where $\hat{U}_s$ is the time evolution operator for a single particle
\begin{align}\label{eq:Us}
\hat{U}_{s}(t) = e^{-i\hat{H}_s t},\quad \hat{H}_s =  \frac{\hat{p}^2}{2m} + \int d^3\mathbf{x}m\Phi(\mathbf{x})\ket{\mathbf{x}}\bra{\mathbf{x}}.
\end{align}
Therefore, at a later time $t$,  the two-particle wavefunction is a symmetrized version of a factorized state. It reads
\begin{align} \label{eq:jhsdgsd}
\ket{\Psi(t)} &\overset{\mathrm{Q1}}{=}\frac{1}{2\sqrt{2}}\left[\hat{U}_s(t)\left(\ket{1L}+\ket{1R}\right)\otimes \hat{U}_s(t)\left(\ket{2L}+\ket{2R}\right)+\hat{U}_s(t)\left(\ket{2L}+\ket{2R}\right)\otimes \hat{U}_s(t)\left(\ket{1L}+\ket{1R}\right)\right]\,.
\end{align}
To quantitatively describe entanglement, we compute again the coefficients $\beta_{ij}$ defined by 
\begin{align}
\ket{\Psi(t)} \overset{Q2}{=} \frac{1}{2}\sum_{ij}\beta_{ij}(t)\ket{1}_{1i}\otimes\ket{1}_{2j}.
\end{align} 
For instance, one of the four coefficients is defined to be $\beta_{LL}\overset{Q2}{=}2\times\prescript{}{1L}{\bra{1}}\prescript{}{2L}{\bra{1}}\hat{U}\ket{\Psi(t_i)}$. Within first quantization, $\beta_{LL}$ reads
\begin{align}\label{eq:aLL}
\beta_{LL}&\overset{Q1}{=}\frac{2}{\sqrt{2}}{\left(\bra{1L}\otimes\bra{2L}+\bra{2L}\otimes\bra{1L}\right)\ket{\Psi(t)}}\nonumber\\
&=\bra{1L}\hat{U}_s(t)\ket{1L}\cdot\bra{2L}\hat{U}_s(t)\ket{2L}+\bra{2L}\hat{U}_s(t)\ket{1L}\cdot\bra{1L}\hat{U}_s(t)\ket{2L}\nonumber\\
&+\bra{1L}\hat{U}_s(t)\ket{1R}\cdot\bra{2L}\hat{U}_s(t)\ket{2L}+\bra{2L}\hat{U}_s(t)\ket{1R}\cdot\bra{1L}\hat{U}_s(t)\ket{2L}\nonumber\\
&+\bra{1L}\hat{U}_s(t)\ket{1L}\cdot\bra{2L}\hat{U}_s(t)\ket{2R}+\bra{2L}\hat{U}_s(t)\ket{1L}\cdot\bra{1L}\hat{U}_s(t)\ket{2R}\nonumber\\
&+\bra{1L}\hat{U}_s(t)\ket{1R}\cdot\bra{2L}\hat{U}_s(t)\ket{2R}+\bra{2L}\hat{U}_s(t)\ket{1R}\cdot\bra{1L}\hat{U}_s(t)\ket{2R}\,.
\end{align}

Similarly, other coefficients are obtained to be
\begin{align}\label{eq:aRL}
\beta_{RL}&\overset{Q1}{=}\frac{2}{\sqrt{2}}{\left(\bra{1R}\otimes\bra{2L}+\bra{2L}\otimes\bra{1R}\right)\ket{\Psi(t)}}\nonumber\\
&=\bra{1R}\hat{U}_s(t)\ket{1R}\cdot\bra{2L}\hat{U}_s(t)\ket{2L}+\bra{2L}\hat{U}_s(t)\ket{1R}\cdot\bra{1R}\hat{U}_s(t)\ket{2L}\nonumber\\
&+\bra{1R}\hat{U}_s(t)\ket{1L}\cdot\bra{2L}\hat{U}_s(t)\ket{2L}+\bra{2L}\hat{U}_s(t)\ket{1L}\cdot\bra{1R}\hat{U}_s(t)\ket{2L}\nonumber\\
&+\bra{1R}\hat{U}_s(t)\ket{1R}\cdot\bra{2L}\hat{U}_s(t)\ket{2R}+\bra{2L}\hat{U}_s(t)\ket{1R}\cdot\bra{1R}\hat{U}_s(t)\ket{2R}\nonumber\\
&+\bra{1R}\hat{U}_s(t)\ket{1L}\cdot\bra{2L}\hat{U}_s(t)\ket{2R}+\bra{2L}\hat{U}_s(t)\ket{1L}\cdot\bra{1R}\hat{U}_s(t)\ket{2R}\,,
\end{align}

\begin{align}\label{eq:aLR}
\beta_{LR}&\overset{Q1}{=}\frac{2}{\sqrt{2}}{\left(\bra{1L}\otimes\bra{2R}+\bra{2R}\otimes\bra{1L}\right)\ket{\Psi(t)}}\nonumber\\
&=\bra{1L}\hat{U}_s(t)\ket{1L}\cdot\bra{2R}\hat{U}_s(t)\ket{2R}+\bra{2R}\hat{U}_s(t)\ket{1L}\cdot\bra{1L}\hat{U}_s(t)\ket{2R}\nonumber\\
&+\bra{1L}\hat{U}_s(t)\ket{1R}\cdot\bra{2R}\hat{U}_s(t)\ket{2R}+\bra{2R}\hat{U}_s(t)\ket{1R}\cdot\bra{1L}\hat{U}_s(t)\ket{2R}\nonumber\\
&+\bra{1L}\hat{U}_s(t)\ket{1L}\cdot\bra{2R}\hat{U}_s(t)\ket{2L}+\bra{2R}\hat{U}_s(t)\ket{1L}\cdot\bra{1L}\hat{U}_s(t)\ket{2L}\nonumber\\
&+\bra{1L}\hat{U}_s(t)\ket{1R}\cdot\bra{2R}\hat{U}_s(t)\ket{2L}+\bra{2R}\hat{U}_s(t)\ket{1R}\cdot\bra{1L}\hat{U}_s(t)\ket{2L}\,,
\end{align}

\begin{align}\label{eq:aRR}
\beta_{RR}&\overset{Q1}{=}\frac{2}{\sqrt{2}}{\left(\bra{1R}\otimes\bra{2R}+\bra{2R}\otimes\bra{1R}\right)\ket{\Psi(t)}}\nonumber\\
&=\bra{1R}\hat{U}_s(t)\ket{1R}\cdot\bra{2R}\hat{U}_s(t)\ket{2R}+\bra{2R}\hat{U}_s(t)\ket{1R}\cdot\bra{1R}\hat{U}_s(t)\ket{2R}\nonumber\\
&+\bra{1R}\hat{U}_s(t)\ket{1L}\cdot\bra{2R}\hat{U}_s(t)\ket{2R}+\bra{2R}\hat{U}_s(t)\ket{1L}\cdot\bra{1R}\hat{U}_s(t)\ket{2R}\nonumber\\
&+\bra{1R}\hat{U}_s(t)\ket{1R}\cdot\bra{2R}\hat{U}_s(t)\ket{2L}+\bra{2R}\hat{U}_s(t)\ket{1R}\cdot\bra{1R}\hat{U}_s(t)\ket{2L}\nonumber\\
&+\bra{1R}\hat{U}_s(t)\ket{1L}\cdot\bra{2R}\hat{U}_s(t)\ket{2L}+\bra{2R}\hat{U}_s(t)\ket{1L}\cdot\bra{1R}\hat{U}_s(t)\ket{2L}\,.
\end{align}

In Ref.~\cite{AzizHowl2025_supp}, as per the orthonormality relations, the authors make the diagonal approximation such that
\begin{align}
\beta^{d}_{LL}\overset{\text{diag}}{\approx} \prescript{}{1L}{\bra{1}}\prescript{}{2L}{\bra{1}}\hat{U}_s(t)\ket{1}_{1L}\ket{1}_{2L}\,.
\end{align}
In first quantization, this amounts to setting the last three lines of Eqs.~\eqref{eq:aLL}-\eqref{eq:aRR} to zero, such that
\begin{align}
\beta^{d}_{LL}=&\,\bra{1L}\hat{U}_s(t)\ket{1L}\cdot\bra{2L}\hat{U}_s(t)\ket{2L}+\bra{2L}\hat{U}_s(t)\ket{1L}\cdot\bra{1L}\hat{U}_s(t)\ket{2L}\,,\label{eq:bll}\\
\beta^d_{RL}=&\,\bra{1R}\hat{U}_s(t)\ket{1R}\cdot\bra{2L}\hat{U}_s(t)\ket{2L}+\bra{2L}\hat{U}_s(t)\ket{1R}\cdot\bra{1R}\hat{U}_s(t)\ket{2L}\,,\label{eq:bRL}\\
\beta^d_{LR}=&\bra{1L}\hat{U}_s(t)\ket{1L}\cdot\bra{2R}\hat{U}_s(t)\ket{2R}+\bra{2R}\hat{U}_s(t)\ket{1L}\cdot\bra{1L}\hat{U}_s(t)\ket{2R}\,,\label{eq:LR}\\
\beta^d_{RR}=&\bra{1R}\hat{U}_s(t)\ket{1R}\cdot\bra{2R}\hat{U}_s(t)\ket{2R}+\bra{2R}\hat{U}_s(t)\ket{1R}\cdot\bra{1R}\hat{U}_s(t)\ket{2R}\,.\label{eq:bRR}
\end{align}
Further, for the discussion concerning entanglement, the authors focus on the transition amplitudes involving the two different objects 1 and 2, and compute the so-called exchange terms. In first quantization, the exchange term for each coefficient, within the diagonal approximation, corresponds to the second term appearing on the right hand side of Eqs.~\eqref{eq:bll}-\eqref{eq:bRR} such that:

\begin{align}
\beta^{d;\text{ex}}_{LL}=&\bra{2L}\hat{U}_s(t)\ket{1L}\cdot\bra{1L}\hat{U}_s(t)\ket{2L},\label{eq:b_ex_ll}\\
\beta^{d;\rm {ex}}_{RL}=&\bra{2L}\hat{U}_s(t)\ket{1R}\cdot\bra{1R}\hat{U}_s(t)\ket{2L}\,,\label{eq:b_ex_RL}\\
\beta^{d;\rm{ex}}_{LR}=&\bra{2R}\hat{U}_s(t)\ket{1L}\cdot\bra{1L}\hat{U}_s(t)\ket{2R}\,,\label{eq:b_ex_LR}\\
\beta^{d;\rm{ex}}_{RR}=&\bra{2R}\hat{U}_s(t)\ket{1R}\cdot\bra{1R}\hat{U}_s(t)\ket{2R}\,.\label{eq:b_ex_RR}
\end{align}
We now present the derivation to obtain the leading order expression for the coefficient $\beta^{d;\rm{ex}}_{LL}$. It can be generalized straightforwardly to obtain the other coefficients. 

At $0^{th}$ order, we have

\begin{align}\label{eq:b_ex_0}
\beta^{d(0);\rm{ex}}_{LL}=\bra{2L}\hat{U}^{(0)}_s(t)\ket{1L}\cdot\bra{1L}\hat{U}^{(0)}_s(t)\ket{2L}=|\braket {2L|1L}|^2 = 0.
\end{align}
This is because $\ket{2L}$ and $\ket{1L}$ are non-overlapping wavepackets as specified in Eq.~\eqref{eq:wavepackets}. 

Then, at $1^{\mathrm{st}}$ order, it follows that also
\begin{align}
\beta^{d(1);\rm{ex}}_{LL}=0,
\end{align}
since 
\begin{align}\label{eq:beta1}
\beta^{d(1);\rm{ex}}_{LL} = U^{(0)1L}_{2L}U^{(1)2L}_{1L}+U^{(1)1L}_{2L}U^{(0)2L}_{1L},
\end{align}
and $U^{(0)1L}_{2L} = U^{(0)2L}_{1L} = 0$. Here, we have defined
\begin{align}
U^{(0)2L}_{1L}:= \bra{2L}\hat{U}^{(0)}_s(t)\ket{1L}.
\end{align}
At $2^{\mathrm{nd}}$ order, due to Eq.~\eqref{eq:b_ex_0}, the only non-trivial contribution can come from the product of two first order transition amplitudes
\begin{align}\label{eq:b_ex_2}
\beta^{d(2);\rm{ex}}_{LL}=\bra{2L}\hat{U}^{(1)}_s(t)\ket{1L}\cdot\bra{1L}\hat{U}^{(1)}_s(t)\ket{2L}.
\end{align}
The transition amplitude $\bra{2L}\hat{U}^{(1)}_s(t)\ket{1L}$, in terms of the propagator, is given by
\begin{align}
\bra{2L}\hat{U}^{(1)}_s(t)\ket{1L} = \frac{m}{V}\int d^3\mathbf{x}_f \int d^3\mathbf{x}_i \int_{t}d^4x \; \theta_{2L}(\mathbf{x_f})G_{0}(t,\mathbf{x}_f;t_x,\mathbf{x})\frac{\Phi(\mathbf{x})}{\hbar}G_{0}(t_x,\mathbf{x};t_i,\mathbf{x}_i)\theta_{1L}(\mathbf{x_i}),
\end{align}
where $G_{0}$ represents  the Schr\"{o}dinger  propagator for a free particle. To model the scenario, as in \cite{AzizHowl2025_supp}, where a freely evolving  wavepacket maintains its shape and does not spread, we must impose
\begin{align}\label{eq:NoSpread}
    \int d^3\mathbf{x}_f\theta_{2L}(\mathbf{x}_f)G_{0}(t,\mathbf{x}_f;t_x,\mathbf{x})\overset{\mathrm{no\,spread}}{=}\theta_{2L}(\mathbf{x}),\quad  \int d^3\mathbf{x}_iG_{0}(t_x,\mathbf{x};t_i,\mathbf{x}_i)\theta_{1L}(\mathbf{x}_i)\overset{\mathrm{no\,spread}}{=}\theta_{1L}(\mathbf{x}).
\end{align}
Doing so, we get 
\begin{align}\label{eq:U1}
\bra{2L}\hat{U}^{(1)}_s(t)\ket{1L} = \frac{m}{V\hbar}\int d^4x \theta_{2L}(\mathbf{x})\Phi(\mathbf{x})\theta_{1L}(\mathbf{x})=0,
\end{align}
since the wavepackets~\eqref{eq:wavepackets} are considered to be non-overlapping,  such that $\theta_{1L}$ and $\theta_{2L}$ cannot both be non-zero at the same location.
Thus, we get 
\begin{align}\label{eq:b_ex_2}
\beta^{d(2);\rm{ex}}_{LL}=0. 
\end{align}
Then, at $3^{\mathrm{rd}}$ order, following the same line of reasoning as in Eq.~\eqref{eq:beta1}, we find that also 
\begin{align}\label{eq:b_ex_3}
\beta^{d(3);\rm{ex}}_{LL}=0.
\end{align}
The first non-vanishing contribution to $\beta^{\rm{ex}}$ can therefore only come at $4^{th}$ order. Using the considerations above, i.e., $\bra{2L}\hat{U}^{(0)}_s(t)\ket{1L} = \bra{2L}\hat{U}^{(1)}_s(t)\ket{1L} = 0$, we see that the only non-vanishing contribution at this order 
is given by
\begin{align}\label{eq:betaLL_Q1}
\beta^{d(4);\rm{ex}}_{LL}=\bra{2L}\hat{U}^{(2)}_s(t)\ket{1L}\cdot\bra{1L}\hat{U}^{(2)}_s(t)\ket{2L}.
\end{align}
In terms of the propagator, we get
\begin{align}\label{eq:betaLL_Q1_propagator}
& \beta^{d(4);\rm{ex}}_{LL}=\nonumber \\
& \left[\frac{m^2}{V}\int d^3\mathbf{x}_f d^3\mathbf{x}_i \int_t d^4x d^4y \;\theta_{2L}(\mathbf{x}_f)G_{0}(t,\mathbf{x}_f;t_x,\mathbf{x})\frac{\Phi(\mathbf{x})}{\hbar}G_{0}(t_x,\mathbf{x};t_y,\mathbf{y})\frac{\Phi(\mathbf{y})}{\hbar}G_{0}(t_y,\mathbf{y};t_i,\mathbf{x}_i)\theta_{1L}(\mathbf{x}_i)\right]^2.
\end{align}
In Ref.~\cite{AzizHowl2025_supp}, the propagator between the two scattering events at $x$ and $y$ is retained, but the free evolution of the asymptotic initial and final wavepackets is neglected as per the no-spread constraint in Eq.~\eqref{eq:NoSpread}. Using the same considerations we get
\begin{align}\label{eq:Scatt_Order2}
\beta^{d(4);\rm{ex}}_{LL}=\left[\frac{m^2}{V}\int_{0}^{t}dt_xdt_y\int d^3\mathbf{x} d^3\mathbf{y} \; \theta_{2L}(\mathbf{x})\frac{\Phi(\mathbf{x})}{\hbar}G_{0}(t_x,\mathbf{x};t_y,\mathbf{y})\frac{\Phi(\mathbf{y})}{\hbar}\theta_{1L}(\mathbf{y})\right]^2.
\end{align}
The expression for the free propagator $G_0$ is well-known. In the instantaneous time limit, as applied in Ref.~\cite{AzizHowl2025_supp}, it is given by
\begin{align}\label{eq:Inst}
    G_{0}(t_x,\mathbf{x};t_y,\mathbf{y}) \rightarrow \;G_{\rm inst}(x-y)
=\delta(t_x-t_y)\,K(\mathbf x-\mathbf y),
\quad
K(\mathbf x-\mathbf y)=K(\mathbf y-\mathbf x) = -\frac{i m}{2\pi\hbar}\,\frac{1}{|\mathbf x-\mathbf y|}.
\end{align}
Using the instantaneous propagator in the expression for the exchange coefficient, we get 
\begin{align}\label{eq:b_4_ll}
\beta^{d(4);\rm{ex}}_{LL}=\left[\frac{im^3t}{2\pi V\hbar^3}\int d^3\mathbf{x} d^3\mathbf{y} \theta_{2L}(\mathbf{x})\frac{\Phi(\mathbf{x})\Phi(\mathbf{y})}{|\mathbf{x}-\mathbf{y}|}\theta_{1L}(\mathbf{y})\right]^2.
\end{align}
The expression above can be straightforwardly generalized to obtain the other exchange coefficients: 
\begin{align}\label{eq:b_4_ij}
\beta^{d(4);\rm{ex}}_{ij}=\left[\frac{im^3t}{2\pi V\hbar^3}\int d^3\mathbf{x} d^3\mathbf{y} \theta_{2j}(\mathbf{x})\frac{\Phi(\mathbf{x})\Phi(\mathbf{y})}{|\mathbf{x}-\mathbf{y}|}\theta_{1i}(\mathbf{y})\right]^2.
\end{align}

The expression above is obtained after discarding the off-diagonal terms, which is why the authors~\cite{AzizHowl2025} find the exchange terms to be non-factorized, even for $N=1$. However, as we showed in Sec.~\ref{subsec:scattering}, for $N=1$, the exchange terms, by themselves, maintain the factorized form of the wavefunction.  Factorization for $N=1$ is in fact restored when the off-diagonals terms are kept, which should not have been discarded in the first place, since the orthonormality relations do not kill the the off-diagonal terms for $N=1$, as can be seen from Eq.~\eqref{eq:orthonormality}.

To see this, it is instructive to review the results from the point of view of dominant contributions to the transition amplitudes. The coefficients $\beta_{ij}^{(4);\rm{ex}}$ scale with the branch distance as in Eq.~\eqref{eq:BranchDistance}. Using the correct expression in Eq.~\eqref{eq:aLL}, one finds instead, for example, that $\beta^{(4);\rm{ex}}_{LL}$ is made of four pieces:
\begin{align}\label{eq:betaLL}
\prescript{}{1L}{\bra{1}}\prescript{}{2L}{\bra{1}}\hat{U}^{(4)}\ket{1}_{1R}\ket{1}_{2L} & =\frac{\kappa}{d_{LL}d_{RL}},\nonumber\\
\prescript{}{1L}{\bra{N}}\prescript{}{2L}{\bra{1}}\hat{U}^{(4)}\ket{1}_{1R}\ket{1}_{2R} & = \frac{\kappa}{d_{LR}d_{RL}},\nonumber\\
\prescript{}{1L}{\bra{1}}\prescript{}{2L}{\bra{1}}\hat{U}^{(4)}\ket{1}_{1L}\ket{1}_{2L} & =\frac{\kappa}{d^2_{LL}},\nonumber\\
\prescript{}{1L}{\bra{N}}\prescript{}{2L}{\bra{1}}\hat{U}^{(4)}\ket{1}_{1L}\ket{1}_{2R}
& = \frac{\kappa}{d_{LL}d_{LR}},
\end{align}
where $\kappa =\left(i 6 G^2m^2M^3Rt\right)^2/\left(25 \hbar^3\right)^2$. The same is true for the other three coefficients. 

The diagonal approximation, which is not valid for $N=1$, keeps only the third line of  Eq.~\eqref{eq:betaLL}, which is, however, {\it subdominant} with respect to the off-diagonal term in the first line, for geometries  where $d_{RL} \ll d_{LL}, d_{LR}, d_{RR}$, as considered in \cite{AzizHowl2025_supp}.   Similar considerations hold for $\beta^{(4);\rm{ex}}_{RR}$ and $\beta^{(4);\rm{ex}}_{LR}$. 

Given this, if one consistently collects the dominant contributions to the exchange terms for each $\beta_{ij}^{(4);\rm{ex}}$,  we get 
\begin{align}\label{eq:dij_offdiag}
    \beta^{(4);\rm{ex}}_{LL}&= \frac{\kappa}{d_{LL}d_{RL}},\quad \beta^{(4);\rm{ex}}_{RR} = \frac{\kappa}{d_{RR}d_{RL}},\nonumber\\
    \beta^{(4);\rm{ex}}_{LR}&=\frac{\kappa}{d_{RR}d_{LL}},\quad  
     \beta^{(4);\rm{ex}}_{RL} 
     = \frac{\kappa}{d^2_{RL}}\,.    
\end{align}
One sees that, indeed, the $RL$ coefficient is dominant with respect to the other three; yet, the overall hierarchy of dependencies is such that also the exchange contribution has a factorized structure
\begin{align}\label{eq:Factorized}
\ket{\Psi^{[4]}_{\rm{ex}}(t)}|_{N=1}
& = \frac{1}{\mathcal N}\left(\frac{1}{d_{LL}}\ket{1}_{1L} + \frac{1}{d_{RL}}\ket{1}_{1R}\right)\otimes\left(\frac{1}{d_{RL}}\ket{1}_{2L}+\frac{1}{d_{RR}}\ket{1}_{2R}\right),
\end{align}
where $\mathcal N$ is a normalization factor.

The calculation shows that the exchange coefficients obtained in Ref.~\cite{AzizHowl2025_supp} can also be derived by staying within first quantization, for non-relativistic identical particles interacting with an external classical potential. For this reason, the physical interpretation of ascribing the exchange terms to virtual particles, as done by the authors, does not hold, since virtual particles do not arise in the framework of the standard non-relativistic Schr\"{o}dinger equation.

\subsection{Distinguishable particles}\label{subsec:Distinguishable}
We compute one more time the coefficients $\beta _{ij}$, within first quantization, but this time for non-relativistic \textit{distinguishable particles}. The initial state reads
\begin{align}\label{eq:InitialQ1Dist}
\ket{\Psi(t_i)} \overset{\mathrm{Q1}}{=} \frac{1}{2}\left(\ket{1L}+\ket{1R}\right)\otimes\left(\ket{2L}+\ket{2R}\right)\,,
\end{align}
The coefficients $\beta_{ij}$, at a later time $t$, are given by:
\begin{align}\label{eq:aLLDist}
\beta_{LL}\overset{Q1}{=}{\bra{1L}\otimes\bra{2L}\ket{\Psi(t)}}
&=\frac{1}{2}\bra{1L}\hat{U}_s(t)\ket{1L}\cdot\bra{2L}\hat{U}_s(t)\ket{2L}+\frac{1}{2}\bra{1L}\hat{U}_s(t)\ket{1R}\cdot\bra{2L}\hat{U}_s(t)\ket{2L}\nonumber\\
&\quad+\frac{1}{2}\bra{1L}\hat{U}_s(t)\ket{1L}\cdot\bra{2L}\hat{U}_s(t)\ket{2R}+\frac{1}{2}\bra{1L}\hat{U}_s(t)\ket{1R}\cdot\bra{2L}\hat{U}_s(t)\ket{2R}\,.
\end{align}
\begin{align}\label{eq:aRLDist}
\beta_{RL}\overset{Q1}{=}\bra{1R}\otimes\bra{2L}\ket{\Psi(t)}
&=\frac{1}{2}\bra{1R}\hat{U}_s(t)\ket{1R}\cdot\bra{2L}\hat{U}_s(t)\ket{2L}+\frac{1}{2}\bra{1R}\hat{U}_s(t)\ket{1L}\cdot\bra{2L}\hat{U}_s(t)\ket{2L}\nonumber\\
&\quad+\frac{1}{2}\bra{1R}\hat{U}_s(t)\ket{1R}\cdot\bra{2L}\hat{U}_s(t)\ket{2R}+\frac{1}{2}\bra{1R}\hat{U}_s(t)\ket{1L}\cdot\bra{2L}\hat{U}_s(t)\ket{2R}\,.
\end{align}

\begin{align}\label{eq:aLRDist}
\beta_{LR}\overset{Q1}{=}\bra{1L}\otimes\bra{2R}\ket{\Psi(t)}
&=\frac{1}{2}\bra{1L}\hat{U}_s(t)\ket{1L}\cdot\bra{2R}\hat{U}_s(t)\ket{2R}+\frac{1}{2}\bra{1L}\hat{U}_s(t)\ket{1R}\cdot\bra{2R}\hat{U}_s(t)\ket{2R}\nonumber\\
&\quad+\frac{1}{2}\bra{1L}\hat{U}_s(t)\ket{1L}\cdot\bra{2R}\hat{U}_s(t)\ket{2L}+\frac{1}{2}\bra{1L}\hat{U}_s(t)\ket{1R}\cdot\bra{2R}\hat{U}_s(t)\ket{2L}\,.
\end{align}

\begin{align}\label{eq:aRRDist}
\beta_{RR}\overset{Q1}{=}\bra{1R}\otimes\bra{2R}\ket{\Psi(t)}
&=\frac{1}{2}\bra{1R}\hat{U}_s(t)\ket{1R}\cdot\bra{2R}\hat{U}_s(t)\ket{2R}+\frac{1}{2}\bra{1R}\hat{U}_s(t)\ket{1L}\cdot\bra{2R}\hat{U}_s(t)\ket{2R}\nonumber\\
&\quad+\frac{1}{2}\bra{1R}\hat{U}_s(t)\ket{1R}\cdot\bra{2R}\hat{U}_s(t)\ket{2L}+\frac{1}{2}\bra{1R}\hat{U}_s(t)\ket{1L}\cdot\bra{2R}\hat{U}_s(t)\ket{2L}\,.
\end{align}
We see that for distinguishable particles, there are no transition amplitudes of the type $\bra{1i}\hat{U}_s\ket{2j}$ ($i,j\in \{L,R\}$) involving the two particles, and that the exchange terms are exactly zero
\begin{align}\label{eq:ExchangeZero}
    \beta^{\rm{ex}}_{ij}(t)= 0.
\end{align}
The direct terms do not lead to entanglement, and the final state for non-relativistic distinguishable particles remains factorized at \textit{all orders}. Indeed, it can be written as 
\begin{align}\label{eq:fact_dist}
\ket{\Psi(t)}= \frac{1}{\mathcal N}\left[\left(U^{1L}_{1L}+U^{1L}_{1R}\right)\ket{1L}+\left(U^{1R}_{1L}+U^{1R}_{1R}\right)\ket{1R}\right]
\otimes\left[\left(U^{2L}_{2L}+U^{2L}_{2R}\right)\ket{2L}+\left(U^{2R}_{2L}+U^{2R}_{2R}\right)\ket{2R}\right],
\end{align}
where $U^{1R}_{1L}:= \bra{1R}\hat{U}_s(t)\ket{1L}$.

\section{Entanglement generation by free evolution: lattice analogue}\label{sec:Lattice}

We present the calculational details of the example discussed in the min text. As outlined there, consider a one-dimensional lattice with five sites, labeled by $i=1,\ldots,5$. Let $\hat a_i^\dagger$ and $\hat a_i$ be bosonic creation and annihilation operators satisfying the canonical commutation relations. At
$t=0$, one boson is prepared in a superposition of sites $1,2$, which define the left region $L$, while the other is prepared in a superposition of sites $4,5$, which define the right region $R$. The initial state is then:
\begin{equation}
\ket{\Psi(0)} = \frac{1}{2}
(\hat a_1^\dagger+\hat a_2^\dagger)
(\hat a_4^\dagger+\hat a_5^\dagger)|0\rangle,
\end{equation}
and, it is clearly factorized.

Over time, the free evolution allows the two particles to hop between nearest-neighbor sites, providing the lattice analogue of diffusion in continuous space. The second-quantized Hamiltonian describing the process is
\begin{equation}
\hat H = J\sum_{n=1}^{4}
\left(
\hat a_{n+1}^\dagger \hat a_n
+
\hat a_n^\dagger \hat a_{n+1}
\right),
\end{equation}
where $J$ is the hopping amplitude. We set $\hbar=1$ and introduce the dimensionless parameter $\epsilon=Jt$.
The time evolution operator is $\hat U(t)=e^{-i\hat Ht}$. In the weak-diffusion regime $\epsilon\ll1$  the dynamics  can be solved perturbatively.

The evolved state at time $t$ is
\begin{equation}
\ket{\Psi(t)}
=
\hat U(t) \ket{\Psi(0)} =
\frac{1}{2} \hat U(t) \left(\hat a_1^\dagger+\hat a_2^\dagger\right) \hat U^\dagger(t)
\hat U(t) \left(\hat a_4^\dagger+\hat a_5^\dagger\right) \hat U^\dagger(t) |0\rangle,
\label{eq:evolved_state}
\end{equation}
where we have used the fact that $\hat{U}^{\dagger}\ket{0} = \ket{0}$. To compute the time evolution, we use the standard expansion
\begin{equation}
\hat U(t) \hat O \hat U^\dagger(t)
=
\hat O-it[\hat H,\hat O]
+\frac{(-it)^2}{2!}[\hat H,[\hat H,\hat O]]
+\cdots ,
\end{equation}
valid for any operator $\hat O$. One has
\begin{equation}
[\hat H,\hat a_j^\dagger]
=
J\left(
\hat a_{j-1}^\dagger+\hat a_{j+1}^\dagger
\right),
\end{equation}
where we have defined $\hat a_0^\dagger=\hat a_6^\dagger=0$. Therefore, each additional commutator with $\hat H$ generates one further nearest-neighbour hop. Explicitly, we have
\begin{align}
\frac{1}{J} \left[ \hat H, \hat a_1^\dagger + \hat a_2^\dagger \right] &=
\hat a_1^\dagger + \hat a_2^\dagger + \hat a_3^\dagger, \nonumber \\
\frac{1}{J^2}
\left[ \hat H, \left[ \hat H, \hat a_1^\dagger + \hat a_2^\dagger \right] \right] & = \hat a_1^\dagger +2 \hat a_2^\dagger + \hat a_3^\dagger + \hat a_4^\dagger, \nonumber \\
\frac{1}{J^3} \left[ \hat H, \left[ \hat H, \left[ \hat H, \hat a_1^\dagger + \hat a_2^\dagger \right] \right] \right]  & = 2\hat a_1^\dagger +2 \hat a_2^\dagger +3 \hat a_3^\dagger + \hat a_4^\dagger + \hat a_5^\dagger, \nonumber \\
\frac{1}{J^4} \left[ \hat H, \left[ \hat H, \left[ \hat H, \left[ \hat H, \hat a_1^\dagger + \hat a_2^\dagger \right] \right] \right] \right] & =
2\hat a_1^\dagger + 5 \hat a_2^\dagger +3 \hat a_3^\dagger + 4\hat a_4^\dagger +\hat a_5^\dagger.
\end{align}
Using $\epsilon=Jt$, we obtain
\begin{equation}\label{eq:LeftRegionEvolution}
\hat U(t) \left( \hat a_1^\dagger + \hat a_2^\dagger\right) \hat U^{\dagger}(t) = \sum_{i=1}^{5}\ell_i \hat a_i^\dagger + \mathcal{O}(\epsilon^5),
\end{equation}
with the coefficients, through fourth order,
\begin{align}
\ell_1
&=
1-i\epsilon
-\frac{\epsilon^2}{2}
+\frac{i\epsilon^3}{3}
+\frac{\epsilon^4}{12},
\\
\ell_2
&=
1-i\epsilon
-\epsilon^2
+\frac{i\epsilon^3}{3}
+\frac{5\epsilon^4}{24},
\\
\ell_3
&=
-i\epsilon
-\frac{\epsilon^2}{2}
+\frac{i\epsilon^3}{2}
+\frac{\epsilon^4}{8},
\\
\ell_4
&=
-\frac{\epsilon^2}{2}
+\frac{i\epsilon^3}{6}
+\frac{\epsilon^4}{6},
\\
\ell_5
&=
\frac{i\epsilon^3}{6}
+\frac{\epsilon^4}{24}.
\end{align}
As we can see, the leading order contributions to the various coefficients are $ \ell_1,\ell_2=\mathcal{O}(1)$, $\ell_3=\mathcal{O}(\epsilon)$, $\ell_4=\mathcal{O}(\epsilon^2)$, $\ell_5=\mathcal{O}(\epsilon^3)$ and higher number of hoppings appear at higher orders in the perturbative parameter $\epsilon$. Thus, at small times ($t\lesssim \mathcal{O}(1/J)$), the particles mainly remain in their original locations and only weakly diffuse elsewhere. 

By reflection symmetry of the chain,
\(1\leftrightarrow5\), \(2\leftrightarrow4\), \(3\leftrightarrow3\), we can compute
\begin{equation}
    \hat U(t)
    \left(\hat a_4^\dagger+\hat a_5^\dagger\right)
    \hat U^\dagger(t)
    =
    \sum_{i=1}^{5}\ell_{6-i}\hat a_i^\dagger
    +
    \mathcal{O}(\epsilon^5),
    \label{eq:right_mode_evolution}
\end{equation}
analogously to the computation presented for Eq.~\eqref{eq:LeftRegionEvolution}. Doing so, we obtain the result for the full evolved state to be
\begin{equation}
\ket{\Psi(t)} = \frac{1}{2}\Big(\sum_{i=1}^{5}\ell_i a_i^\dagger\Big)
\Big(\sum_{j=1}^{5}\ell_{6-j} a_j^\dagger
\Big)|0\rangle + \mathcal{O}(\epsilon^5).
\end{equation}
The above equation shows directly that the full state remains factorized with respect to the partition defined by
the freely evolved modes. That is, we can define
\begin{equation}
    \hat a_L^\dagger(t)
\equiv
\hat U(t)\left(\hat a_1^\dagger+\hat a_2^\dagger\right)
\hat U^\dagger(t),
\qquad
\hat a_R^\dagger(t)
\equiv
\hat U(t)\left(\hat a_4^\dagger+\hat a_5^\dagger\right)
\hat U^\dagger(t),
    \label{eq:evolved_modes}
\end{equation}
such that
\begin{equation}
    \ket{\Psi(t)}
    =
    \frac{1}{2}\hat a_L^\dagger(t)\hat a_R^\dagger(t)\ket{0}.
    \label{eq:factorized_evolved_modes}
\end{equation}
The factorization of Eq.~\eqref{eq:factorized_evolved_modes} reflects the fact that the two particles do not interact and evolve independently.

We now mimic the projection considered in \cite{AzizHowl2025_supp}. Namely, we retain only the sector in which
exactly one boson is found in $L=(1,2)$ and exactly one boson is found in $R=(4,5)$.
The corresponding unnormalised projected state is
\begin{align}
\ket{\tilde\Psi(t)}
= \frac{1}{2}\left[
\beta_{14} \hat a_1^\dagger \hat a_4^\dagger|0\rangle
+
\beta_{15} \hat a_1^\dagger \hat a_5^\dagger|0\rangle +
\beta_{24} \hat a_2^\dagger \hat a_4^\dagger|0\rangle
+
\beta_{25} \hat a_2^\dagger \hat a_5^\dagger|0\rangle \right],
\end{align}
where
\begin{align}
\beta_{14} & =  \ell_1\ell_2+\ell_5\ell_4, \qquad
\beta_{15}  =  \ell_1^2+\ell_5^2, 
\nonumber \\
\beta_{24} & =   \ell_2^2+\ell_4^2, \qquad
\beta_{25} =  \ell_2\ell_1+\ell_4\ell_5. 
\label{eq:coeff}
\end{align}
The coefficient matrix for the post-selected state is therefore
\begin{equation}
\beta(\epsilon) = 
\begin{pmatrix}\label{eq:Cmatrix}
\beta_{14}&\beta_{15}\\
\beta_{24}&\beta_{25}
\end{pmatrix},
\end{equation}
whose determinant is
\begin{equation}
\det \beta 
= 
- 
( \ell_1\ell_4 - \ell_2\ell_5)^2 
= 
-\frac{\epsilon^4}{4} 
+{\mathcal O}(\epsilon^5).
\label{eq:determinant}
\end{equation}
Thus, for any small but nonzero $\epsilon$, the matrix $\beta(\epsilon)$ has rank two, meaning that
the coefficients cannot be factorized as $\beta_{ij}(\epsilon)=A_i(\epsilon)B_j(\epsilon)$.
The state obtained after projecting onto the sector with one boson in $L$ and one in $R$ is
therefore entangled with respect to the $L|R$ partition. In particular, it is important to notice that no interaction or potential has been introduced: the rank-two structure already appears for freely
evolving identical bosons.

The origin of this rank-two structure can be made explicit by rewriting the evolved modes as
\begin{align}\label{eq:evolvedModes_DPlusE}
\hat a_L^\dagger(t) 
&= \left({a}_1\hat{a}^\dagger_1+{a}_2\hat{a}^\dagger_2\right) + \left(c_1\hat{a}^\dagger_4+c_2\hat{a}^\dagger_5\right),
\nonumber\\
\hat a_R^\dagger(t) 
&=\left({b}_1\hat{a}^\dagger_1+{b}_2\hat{a}^\dagger_2\right) + \left(d_1\hat{a}^\dagger_4+d_2\hat{a}^\dagger_5\right),
\end{align}
where we have omitted the terms proportional to $\hat a_3^\dagger$, since they do not contribute
to the sector with one particle in $L$ and one in $R$. The vectors appearing in this decomposition are
\begin{equation}
a= \binom{\ell_1}{\ell_2},
\qquad
c=\binom{\ell_4}{\ell_5},
\qquad
b=\binom{\ell_5}{\ell_4},
\qquad
d=\binom{\ell_2}{\ell_1}.
\end{equation}
Here $a,d=\mathcal O(1)$, while $b,c=\mathcal O(\epsilon^2)$.

A configuration with one particle in each region receives contribution from two different types of processes. First, $\hat a_L^\dagger(t)$ may create a particle in $L$ and $\hat a_R^\dagger(t)$ a particle
in $R$. These are the so-called direct terms, which describe the process where particles do not leave the region they were originally created in, at the initial time $t=0$. These processes correspond to the coefficients $a_i$ and $d_i$ in Eq.~\eqref{eq:evolvedModes_DPlusE}  giving the contribution $ad^T=\mathcal O(1)$ to the coefficient matrix $\beta$ in Eq.~\eqref{eq:Cmatrix}. The second type of process is where $\hat a_L^\dagger(t)$ creates a
particle in $R$ (coefficients $c_i$) and $\hat a_R^\dagger(t)$ in $L$ (coefficients $b_i$), giving the so-called exchange contribution
$bc^T=\mathcal O(\epsilon^4)$ to the full coefficient matrix $\beta$ in Eq.~\eqref{eq:Cmatrix}. The exchange contribution is nonzero because the free evolution allows
particles to hop from one region to the other. Therefore, at a later time, in the post-selected sector, we get
\begin{equation}\label{eq:Direct+Exchange}
\beta = ad^T + bc^T.
\end{equation}
Consistently with, and analogously to, the discussion at the end of Sec.~\ref{subsec:ExchangeTerms_QFT}, we see that both the direct and exchange terms contribute to the final state. Moreover, when considered separately, the direct and the exchange terms preserve the factorized form of the statevector. This can be seen from the fact that $\text{det}(ad^T)=0$ and  $\text{det}(bc^T)=0$. However, when considered together, the state remains non-factorized, since for the full coefficient matrix we have 
\begin{equation}
\det \beta 
= 
-\bigl(\ell_1\ell_4-\ell_2\ell_5\bigr)^2,
\end{equation}
which is non-zero, as already specified in Eq.~\eqref{eq:determinant}. 

\end{widetext}

\bibliography{entanglement}

\end{document}